   \documentclass[12pt,preprint]{aastex}

\usepackage{natbib}

\begin{document}

\title{Extracting the size of the cosmic electron-positron anomaly}

\author{Katie Auchettl\altaffilmark{1,2} and 
        Csaba Bal\'azs\altaffilmark{1,2}}
\affil{School of Physics, Monash University, Melbourne, Victoria 3800 Australia}

\altaffiltext{1}{Monash Centre for Astrophysics, Monash University, Victoria 3800 Australia}

\altaffiltext{2}{ARC Centre of Excellence for Particle Physics at the Tera-scale, Monash University, Victoria 3800 Australia}

\begin{abstract}
We isolated the anomalous part of the cosmic electron-positron flux within a Bayesian likelihood analysis.  Using 219 recent cosmic ray spectral data points, we inferred the values of selected cosmic ray propagation parameters.  In the context of the propagation model coded in GalProp, we found a significant tension between the electron positron related and the rest of the fluxes.  Interpreting this tension as the presence of an anomalous component in the electron-positron related data, we calculated background predictions for PAMELA and Fermi-LAT based on the non-electron-positron related fluxes.  We found a deviation between the data and the predicted background even when uncertainties, including systematics, were taken into account.  We identified this deviation with the anomalous electron-positron contribution.  We briefly compared this model independent signal to some theoretical results predicting such an anomaly. 
\end{abstract}


\section{Introduction}\label{sec.intro} 
 
 
Over the last few decades observations of cosmic rays established an increasingly significant and puzzling deviation from theoretical predictions.  Several experiments, such as 
TS \citep{Golden:1992zm}, 
AMS \citep{Alcaraz:2000bf}, 
CAPRICE \citep{Boezio:2001ac}, 
MASS \citep{Grimani:2002yz}, and
HEAT \citep{Barwick:1997ig, Beatty:2004cy}
provided a hint of an excess of high energy positrons 
in our locality.  
Recent measurements of the PAMELA satellite confirmed these suspicions by establishing an excess in the positron fraction over the theoretical predictions for energies above 10 GeV \citep{Adriani:2008zr}.  The PAMELA data appear to significantly deviate from the background predictions even when sizeable experimental and theoretical uncertainties are taken into account \citep{Delahaye:2008ua, Delahaye:2010ji, Mertsch:2010qf, Timur:2011vv}. 

A possible excess in the electron-positron sum was also indicated by 
AMS \citep{Aguilar:2002ad}, 
PPB-BETS \citep{Torii:2008xu}, and 
HESS \citep{Aharonian:2008aa,Aharonian:2009ah}. 
The exceptionally precise measurement of the the local electron+positron flux by the Fermi-LAT satellite, at first glance, seems to partially confirm the electron+positron excess above 100 GeV \citep{Ackermann:2010ij}.  The deviation between the Fermi-LAT data (especially the 2010 release) and the theoretical background calculation, produced by the numerical code GalProp by \citet{Strong:1998pw} appears to be significant.  These results were recently confirmed by the PAMELA collaboration which measured the cosmic ray electron flux in a similar energy range and found it to be consistent with the Fermi-LAT data.

The deviation between the measurements and the predicted backgrounds prompted numerous attempts to explain it by invoking new physics ranging from modification of the cosmic ray propagation 
\citep{Stawarz:2009ig, Cowsik:2009ga, Katz:2009yd, Blasi:2009hv, Hu:2009bc, Dado:2009ux, Perelstein:2010fq, Perelstein:2010gt}, 
through supernova remnants 
\citep{Ahlers:2009ae, Shaviv:2009bu, Fujita:2009wk, Hooper:2008kg, Yuksel:2008rf, Profumo:2008ms, Malyshev:2009tw, Barger:2009yt, Grasso:2009ma, Mertsch:2009ph, Malyshev:2009zh}, 
to dark matter annihilation 
\citep{Cirelli:2008pk, ArkaniHamed:2008qn, Cholis:2008qq, Harnik:2008uu, Allahverdi:2008jm, Calmet:2009uz, Shirai:2009fq, Chen:2009mj, Hamaguchi:2009jb, Okada:2009bz, Fukuoka:2009cu, Bai:2009ka, Shirai:2009wi, Chen:2009zpa, Mardon:2009gw, Demir:2009kc, Hooper:2009gm, Choi:2009qc, Feldman:2009wv, Yin:2008bs, Hamaguchi:2008rv, Ibarra:2008jk, Nardi:2008ix, Ishiwata:2008qy, DeLopeAmigo:2009dc, Arvanitaki:2009yb, Buchmuller:2009xv, Ibarra:2009dr, Chen:2009gd, He:2009ra, Hooper:2008kv, Brun:2009aj,  Feldman:2008xs, Ibe:2008ye, Guo:2009aj, Bi:2009uj, Hisano:2004ds, MarchRussell:2008yu, Dent:2009bv, Zavala:2009mi, Feng:2009hw, Backovic:2009rw, Ciafaloni:2011sa, Zeldovich:1980st, PhysRevD.52.1828,Fargion:1999ss,Belotsky:2004st}.  
\cite{Serpico:2011wg} summarizes the present situation of these speculations.  



The existence and statistical severity of the electron-positron anomaly depends on the theoretical prediction of the cosmic ray background.  While the origin of the cosmic rays is not fully understood their local observation, coupled with other astrophysical measurements, enables us to build and constrain a model of particle production and propagation in our Galaxy.  Such a model is based on the relatively well understood features of particle diffusion within the Milky Way.  The diffusion is described by the transport equation, subject to an initial source distribution and boundary conditions.  The local electron and positron fluxes are calculated by solving this transport equation.  Besides the lack of precise knowledge of the cosmic ray sources, the background prediction is challenging because the propagation model has numerous free parameters, such as the convection velocities, spatial diffusion coefficients, or momentum loss rates.  


Motivated by possible new physics buried in the Fermi-LAT data, in this work we attempt to determine the size of the anomalous contribution in the cosmic electron-positron flux.  Our strategy involves the following main steps.
\begin{itemize}
\item Finding the cosmic ray propagation parameters that influence the electron-positron flux measured by Fermi-LAT and PAMELA the most.  
\item Subjecting cosmic ray data, other than the Fermi-LAT and PAMELA electron-positron measurements to a Bayesian likelihood analysis, to determine the 68 \% (1-$\sigma$) credibility regions of the relevant propagation parameters.
\item Calculating a background prediction, with uncertainties, for Fermi-LAT and PAMELA, based on the determined 1-$\sigma$ credibility regions of the propagation parameters.
\item Subtracting the background prediction from the Fermi-LAT and PAMELA measurement to isolate the anomalous part of the spectrum.
\end{itemize}
Since in the process of the likelihood analysis we determine the uncertainty of the electron-positron background, we can also quantify the statistical significance of the deviation between the cosmic ray data and the theoretical background calculation.  

When contrasted with the earlier literature, our work contains two main novel results.  1. Demonstration of a significant tension between the electron-positron related and the rest of the cosmic ray data in the context of the propagation model coded in GalProp.  2. The extraction of the anomalous part of the electron-positron flux.  We were able to obtain these results because we use more data than other similar studies \citep{Maurin:2001sj, Maurin:2002hw, 2010A&A...516A..67M, 2010A&A...516A..66P, Lin:2010fba, Trotta:2010mx}.  We used the numerical code GalProp in the Bayesian framework, extending the analysis of \cite{Trotta:2010mx} to quantify the uncertainty in the background contribution of cosmic ray spectra.  Unlike \cite{Lin:2010fba} we do not use gamma ray data because some components of the gamma ray flux are thought to be affected by the same (or similar) anomalous contributions as the electron-positron flux.  Leaving out the calculation of gamma ray propagation also speeds up our numerical calculations.  We decided to include gamma ray data in our analysis at a later stage.

While the numerical analysis by \cite{Trotta:2010mx} is very similar to ours the choice of the free diffusion and nuisance parameters are different.  More importantly, the use of substantially more experimental data (219 data points altogether compared to 76 in \cite{Trotta:2010mx}), enables us to constrain the background prediction well enough to isolate the anomalous part of the $e^+ e^-$ flux.  The experimental data we use, come from multiple instruments, over a wide energy range, as discussed in section \ref{sec:ExpDat}.

\section{Cosmic ray propagation}

Cosmic rays are highly energetic particles which have their origins locally and remotely in the visible universe \citep{1964ocr..book.....G, 1987PhR...154....1B, Stawarz:2009ig, Aharonian:2011da}.  They are divided into two main categories: primary and secondary.  Primary cosmic rays are particles that are accelerated by astrophysical objects, such as supernova remnants.  These cosmic rays interact with interstellar matter to create secondary cosmic rays \citep{1987PhR...154....1B, Delahaye:2008ua, Nakamura:2010zzi, Aharonian:2011da}.  
The majority of cosmic ray electrons, for example, are likely to originate from supernova remnants, while cosmic ray positrons are believed to be mainly produced via secondary production processes such as spallation and nucleosynthesis \citep{1987PhR...154....1B, Adriani:2008zr, Delahaye:2009gd, Nakamura:2010zzi, Aharonian:2011da}.

Cosmic ray propagation through the Galaxy is typically quantified using the diffusion model \citep{Ginzburg:1990sk, Schlickeiser:2002pg, Ptuskin:2005ax, Strong:2007nh}.  Diffusion of cosmic rays provides a simple explanation for the highly isotropic distribution of high energy charged particles and their noticeable retention in the Galaxy.  Diffusion results from the particle scattering of cosmic rays on random magnetohydrodynamic (MHD) waves and inhomogeneities in the Galactic magnetic field.  The random nature of the Galactic magnetic field, causes the trajectories of the cosmic rays to become jumbled, causing them to undergo a random walk in space \citep{1964ocr..book.....G, Strong:2007nh, Cotta:2010ej, Aharonian:2011da}.  The energy distribution of cosmic rays are modified by energy losses experienced by these particles as they propagate through the Galaxy.  Energy losses arise due to the interaction of the cosmic rays with the interstellar medium and interstellar radiation fields.  Re-acceleration due to interstellar shocks and Galactic winds powered by convection also contribute \citep{Strong:1998pw, Strong:2007nh, Fan:2010yq}, while for heavy and unstable nuclei, fragmentation processes also need to be taken into account.  

The diffusion model assumes homogeneous propagation of charged particles within the Galactic disk (similar to one of the simplest models of propagation called the leaky box model) but it also takes into account cooling effects.  The density $\psi (\vec r,p,t)$ (per unit particle momentum $p$) of a particular cosmic ray species at a Galactic radius of $\vec r$ can be calculated solving the cosmic ray transport equation which has the general form \citep{Strong:2007nh}
\newcommand{\Dpp}{D_{pp}}
\newcommand{\Dxx}{D_{xx}}
\newcommand{\ddp}{{\partial\over\partial p}}
\begin{eqnarray}
\label{eq:transport}
{\partial \psi (\vec r,p,t) \over \partial t} 
&= &
q(\vec r, p, t) 
+ \vec\nabla \cdot ( \Dxx\vec\nabla\psi - \vec V\psi ) \nonumber \\
& +& \ddp\, \left( p^2 \Dpp \ddp\, {1\over p^2}\, \psi \right)  
- {\partial\over\partial p} \left( \dot{p} \psi
- {p\over 3} \, (\vec\nabla \cdot \vec V )\psi \right)
- {1\over\tau_f}\psi - {1\over\tau_r}\psi\ .
\end{eqnarray}
Here $q(\vec r, p, t)$ is the source term which depends on the production mechanism of primary and secondary cosmic ray contributions.
The spatial diffusion coefficient $D_{xx}$ describes the scattering of cosmic ray species through turbulent magnetic fields.  This propagation can be isotropic or anisotropic, and can be influenced by the cosmic rays themselves \citep{Strong:2007nh}.  Generally, $D_{xx}$ has the form
\begin{eqnarray}
\label{eq:Dxx}
D_{xx} =  D_{0xx} \beta \left({R\over {\rm GeV}}\right)^\delta ,
\end{eqnarray}
where $\beta = v/c$, and
$R=pc/eZ$ is the magnetic rigidity of the particles which describes a particle's resistance to deflection by a magnetic field.  Here $Z$ is the effective nuclear charge of the particle, $v$ is its velocity, $p$ is its momentum, $e$ is its charge, and $c$ is the speed of light.  The energy of high momentum cosmic ray electrons and positrons, for example, can be approximated by $E \simeq eR$ \citep{1984ARA&A..22..425H}.  
The constant exponent $\delta$ indicates the power law dependence of the spatial diffusion coefficient $D_{xx}$.  Different regions of the energy spectra can have different $\delta$ values, producing a discontinuity in the derivative of $D_{xx}$.  This artificial kink is introduced so that one can fit the B/C ratio data over all energies \citep{Strong:1998pw}.

In Eq.(\ref{eq:transport}), $\vec V$ describes the convection velocity which is a function of Galactic radius $r$ and depends on the characteristics of the Galactic winds.  The convection velocity is assumed to have a cylindrical symmetry and increase linearly with height $z$ from the Galactic plane.  Apart from transporting particles through the Galaxy, convection also causes the adiabatic energy losses (or gains) of cosmic rays due to their interaction with the non-uniform flow of gas (Galactic winds) with an inhomogeneous magnetic field.  This is represented by the term $\vec\nabla \cdot \vec V$.

Diffusion in momentum space (diffusive re-acceleration) is described by the coefficient $\Dpp$.  This arises from the scattering of cosmic ray particles on randomly moving MHD waves.  Diffusion in momentum is related to spatial diffusion via
\begin{eqnarray}
\label{eq:DppDxx}
\Dpp\Dxx = {4 p^2 {v_A}^2\over 3\delta(4-\delta^2)(4-\delta) w}\ .
\end{eqnarray}
Here $v_A$ is the Alfven speed, the parameter $w$ characterises the level of hydromagnetic turbulence experienced by the cosmic rays in the interstellar medium \citep{1994ApJ...431..705S}.  This parameter is also known as the ratio of MHD energy density to the magnetic field energy density. 
In the last two terms of Eq.(\ref{eq:transport}) the parameter $\tau_f$ is the time-scale of the fragmentation loss, and $\tau_r$ is the radioactive decay time-scale. 

Observations of galaxies other than ours suggest that cosmic rays are diffusing in a cylindrical slab, whose height is dependent on the Galaxy itself \citep{Delahaye:2009gd}. Consequently, the transport equation (generalised and simplified) is solved in a diffusive region shaped as a solid flat cylinder.  This cylinder, parametrised by the 
coordinates $(\vec r, \phi, z)$, encloses the Galactic plane with height $2 L$ in the $z$-direction ($z$ $\in$ $[-L, L]$) and a radius of $R = 20$ kpc in the $\vec r$ direction.  The solar system is located at $(\vec r, \phi, z)$= (8.5 kpc, 0, 0), while the boundary conditions imposed on this scenario allow the cosmic ray density to vanish at the surface of the flat cylinder and particles may propagate freely outside it and escape.  The rate of energy loss, $b(E)$ is determined by the photon density, strength of the magnetic field and the Thomson scattering cross section associated with the cosmic rays. 

To obtain an explicit analytic solution for a particular cosmic ray species, it is possible to solve the simplified version of the transport equation, such as
\begin{eqnarray}
\label{eq:etrans}
\frac{\partial \psi (E,\vec{r})}{\partial t}
&= &
 D_{xx}\nabla^2 \psi (E,\vec{r})
 +
 \frac{\partial}{\partial E}
 \left[b(E)\psi (E,\vec{r})\right]%
 +
 q(E,\vec{r}),
\end{eqnarray}
for electrons \citep{Delahaye:2007fr}, using a Green's function method \citep{PhysRevD.59.023511}.  However in most cases, that require a realistic description of the astrophysical environment which produce the experimentally observed cosmic ray spectra, an analytical solution is not possible.  Hence a numerical solution is pursued. 

The numerical Galactic cosmic ray propagation package GalProp calculates the propagation of relativistic charged particles and their diffuse emission produced during their propagation through the Galaxy.  GalProp solves the propagation equation numerically for $Z \geq 1$ nuclei, as well as for electrons and positrons on a two dimensional spatial grid with cylindrical symmetry in the Galaxy \citep{Strong:2007nh}.  It also has the capability of solving the diffusion equation in three dimensions.  GalProp starts with the heaviest primary element defined by the user and the propagated solution is used to compute the source term for the secondary products of this element.  This process is continued until protons, secondary electrons, positrons and anti-protons are produced and a steady state solution is obtained.  The cosmic ray spectrum is used to compute the gamma rays and energy losses such as synchrotron radiation experienced by the cosmic rays.  These are computed in conjunction with realistic maps of the interstellar gas distributions and radiation fields based on current HI and CO surveys and detailed theoretical calculations of the Galactic magnetic field \citep{Strong:2007nh}.

The input parameter file for GalProp has a number of free parameters which are available for the author to define.  The main free parameters determine the geometry of the model (radius, height of cylinder, and grid spacing), the distribution of cosmic ray sources (which is usually chosen to represent an even distribution of supernova remnants), the primary cosmic ray spectral shape and the isotropic composition of the sources \citep{Strong:2007nh}, the spatial and momentum diffusion coefficients and their dependence on the particle rigidity \citep{Grasso:2009ma}.
These can be classified into a number of subsets: the diffusion of cosmic ray, the primary cosmic ray sources and radiative energy losses of these primary cosmic rays.  The diffusion subset is described by the parameters defined above: 
\begin{eqnarray}
 D_{0xx}, \delta, L, v_A, \partial {\vec V} / \partial z . 
\end{eqnarray}
The most relevant parameters in the primary cosmic ray source subset are: 
\begin{eqnarray}
 R_{ref}^{e^-}, \gamma^{e^-}, R_{ref}^{nucleus}, \gamma^{nucleus} .
\end{eqnarray}
Here $\gamma^{e^-}$ is the primary source electron injection index.  This specifies the steepness of the electron injection spectrum, $dq(p)/dp \propto p^{\gamma^{e^-}}$, below a reference rigidity $R_{ref}^{e^-}$.  There is also a separate injection index for nuclei defined by $\gamma^{nucleus}$ above $R_{ref}^{nucleus}$.  For further details we refer the reader to \cite{Strong:2007nh}.  

\section{Bayesian inference}

Our aim is to isolate the anomalous part of the Fermi-LAT and PAMELA electron-positron fluxes.  To do this first we need to know the non-anomalous, standard astrophysical background contributing to Fermi and PAMELA.  To extract this background and to determine its uncertainty we use the cosmic ray measurements which appear to be consistent with the background estimates.  First we determine the values of the Galactic propagation parameters most favored by this part of the data and extract the uncertainties of these parameters.  Then we use these parameter values to calculate the $e^-$ and $e^+$ background and its uncertainty and compare this to the measurements of Fermi and PAMELA.  This way we are able to isolate the size and uncertainty of the contribution of the (possible) new source(s) in the electron-positron related fluxes.

To extract the values of the propagation parameters $P = \{p_1, ..., p_N \}$ favored by the experimental data $D = \{d_1, ..., d_M \}$ we utilize Bayesian inference.  In the Bayesian framework the probability density of a certain theoretical parameter $p_i$ acquiring a given value is given by the marginalized posterior probability distribution 
\begin{eqnarray}
 \mathcal{P}(p_i|D) = \int \mathcal{P}(P|D) \prod_{i \ne j=1}^N dp_j ,
 \label{eq:margin1}
\end{eqnarray}
where the integral is carried out over the full range of the parameters.  According to Bayes' theorem the posterior probability density over the full parameter space is calculated as
\begin{eqnarray}
 \mathcal{P}(P|D) = \mathcal{L}(D|P) \frac{\mathcal{P}(P)}{\mathcal{E}(D)} .
\end{eqnarray}
Here the likelihood function $\mathcal{L}(D|P)$ is the conditional probability density of the theoretical predictions for the data with given parameter values $P$.  Data independent information on the parameter distribution is folded in via the prior distribution $\mathcal{P}(P)$ and the Bayesian evidence $\mathcal{E}(D)$, for our purposes, acts as a normalization factor. 

The likelihood function, in our case, is calculated as 
\begin{eqnarray}
 \label{eq:L}
 {\cal L}(D|P) = \prod_{i=1}^M \frac{1}{\sqrt{2 \pi} \sigma_i} \exp(-\chi_i^2(D,P)/2) ,
\end{eqnarray}
where
\begin{eqnarray}
 \label{eq:chi2}
 \chi_i^2(D,P) = \left(\frac{d_i - t_i(P)}{\sigma_i}\right)^2 .
\end{eqnarray}
The log-likelihood $\chi_i^2$ contrasts the central value of the $i$-th data point $d_i$ with the theoretical prediction $t_i$ for given parameter values $P$, in terms of the combined theoretical and experimental uncertainty $\sigma_i$.  

For parameter estimation the Bayesian evidence only plays the role of an irrelevant normalization.  Nevertheless, it is useful to calculate $\mathcal{E}(D)$ when assessing the validity of the hypothesis quantified as the theory parametrized by $P$.  The evidence is easily calculated using the normalization of the posterior density 
\begin{eqnarray} 
 \int \mathcal{P}(P|D) \prod_{j=1}^N dp_j = 1 .
 \label{eq:E}
\end{eqnarray}
This enables us to recast Bayes' theorem in the integral form 
\begin{eqnarray}
 \mathcal{E}(D) = \int \mathcal{L}(D|P) \mathcal{P}(P) \prod_{j=1}^N dp_j .
\end{eqnarray}

Once the posterior distribution is known we can determine the credibility intervals for each of the parameters.  We define a credibility region ${\mathcal R}_x$ for parameter $p_i$ by the collection of minimal sized parameter regions supporting $x$ \% of the total probability:
\begin{eqnarray}
 x = \int_{{\mathcal R}_x} {\mathcal P}(p_i|D) \, dp_i .
\end{eqnarray}
In plain terms, a 68 \% credibility interval is the minimal parameter region that contains 68\% of the area under the posterior distribution.   This region gives the value of the parameter at 1-$\sigma$ certainty.
Combined credibility regions over multi-dimensions of the parameter space can be similarly defined as the minimal region satisfying 
\begin{eqnarray}
 x = \int_{{\mathcal R}_x} {\mathcal P}(p_i,p_j|D) \, dp_i \, dp_j.
\end{eqnarray}

\subsection{Parameter choice}

The calculation of the posterior probability distributions $\mathcal{P}(p_i|D)$ requires us to numerically integrate over the parameter region where the (cumulative) likelihood function is non-negligible.  The CPU demand to reliably sample the posterior density depends on the number of free theoretical parameters $N$ and the speed of the numerical implementation.  In the case of the diffusion model encoded in GalProp the number of free input parameters is around a hundred and for a given set of parameters the code runs for several minutes on a single CPU.  This makes it unfeasible to attempt the calculation of the posterior without simplifications. 

Fortunately, both the number of relevant free parameters and the running time can be substantially reduced.  To reduce the dimension of the parameter space we tested the robustness of the electron-positron flux against the variation of nearly all individual parameters and found that it is mostly sensitive to the following propagation parameters:
\begin{eqnarray} 
\label{eq:P}
P = \{\gamma^{e^-}, \gamma^{nucleus}, \delta_1, \delta_2, D_{0xx} \} .
\end{eqnarray}
Here $\gamma^{e^-}$ and $\gamma^{nucleus}$ are the primary electron and nucleus injection indices parameterizing an injection spectrum without a break, $\delta_1$ and $\delta_2$ are spatial diffusion coefficients below and above a reference rigidity $\rho_0$, and $D_{0xx}$ determines the normalization of the spatial diffusion coefficient.

We found that the electron-positron spectra are fairly insensitive to the rest of the parameters.  We also found that the electron-positron spectrum is not only sensitive to the power law dependence of the spatial diffusion coefficient $D_{xx}$, but the presence of a kink therein.  So following \cite{Strong:2004de} we introduce two coefficients $\delta_1$ and $\delta_2$ which parametrize this power law dependence in Eq.(\ref{eq:Dxx}) below and above a reference rigidity.  We fix this reference rigidity to 4 GeV as \cite{Strong:2004de}.


Our calculations also confirmed the findings of a recent study by \cite{Cotta:2010ej} that the electron-positron flux is sensitive to the change of the Galactic plane height $L$.  Indeed \cite{1994ApJ...431..705S} have shown that there is a connection between $L$ and $D_{0xx}$: 
\begin{eqnarray} 
\label{eq:D0xx}
D_{0xx} = \frac{2 c (1-\delta) L^{1-\delta}}{3 \pi w \delta (\delta + 2)} .
\end{eqnarray}
Thus, varying the cylinder height amounts to the redefinition of $D_{0xx}$ as also noticed by Ref. \cite{DiBernardo:2010is}.  In the light of this, we fix $L$ to 4 kpc and use $D_{0xx}$ as free parameter. 



We treat the normalizations of the $e^-$, $e^+$, ${\bar {\rm p}}$/p, B/C, (SC+Ti+V)/Fe and Be-10/Be-9 fluxes as theoretical nuisances parameters.  
\begin{eqnarray} 
\label{eq:Pn}
P_{nuisance} = \{ \Phi_{e^-}^0, \Phi_{e^+}^0, \Phi_{{\bar p}/p}^0, \Phi_{B/C}^0, \Phi_{(SC+Ti+V)/Fe}^0, \Phi_{Be-10/Be-9}^0 \} .
\end{eqnarray}
They are kept free because the electron-positron flux is either directly or indirectly sensitive to these parameters.  On the other hand, prior information is available for these parameters enabling us to reduce them to the nuisance level.  Since GalProp calculates normalizations based on local cosmic ray measurements, the results of this calculation can be used as a guideline to the central values of the nuisance parameters.  The uncertainties of the normalizations can be reliably estimated by an initial scan over the full parameter space.\footnote{During our analysis of $e^\pm$ related or other data we found that the posterior for $\Phi_{e^\pm}^0$ prefers about 10 \% lower normalization than the value GalProp determines.  Since these normalizations form part of our parameters in our plots we use the posterior normalizations rather than the GalProp ones.}

Varying the parameters listed in Eq.(\ref{eq:P}) and (\ref{eq:Pn}), we confirmed the result of \cite{Trotta:2010mx} that the electron+positron flux of Fermi-LAT can be well reproduced by the theoretical calculation.  We also found that by changing these parameters the theory can match well the latest PAMELA electron spectrum \citep{Adriani:2011xv} and the latest PAMELA positron fraction data \citep{Adriani:2010ib}.  (We defer the discussion of the quantitative details to the results section.)  This demonstrates that varying the selected parameters gives us enough flexibility to fit all the observed features of the electron-positron spectra.  


While the Galactic propagation of GeV or higher energy cosmic rays is relatively well understood, the propagation of a few GeV or lower energy electrons and positrons in the turbulent, magnetized interstellar medium, remains a formidable challenge \citep{Prantzos:2010wi}.  Local effects, such as solar modulation and the geomagnetic cutoff, significantly affect cosmic rays at lower energies \citep{PesceRollins:2009tu}.  Since solar modulation effects, based on the force field model, are built into GalProp, we take these effects into account by varying the value of the modulation potential in the code.  Following \cite{Gast:2009}, we assume a charge-sign dependent modulation, that is positively and negatively charged cosmic rays are modulated differently by the Sun.  \cite{Gast:2009} conclude that the effect of this charge dependent modulation on (PAMELA) positrons is substantial.  They also show that the modulation effect on the ${\bar p}/p$ ratio is comparable to the statistical uncertainties.  As described in the next section, we absorb this effect in the systematic uncertainties of the ${\bar p}/p$ data.  Heavier nuclei (B, C, Sc, Ti, V, Fe and Be) can carry higher positive charges than that of the proton, but their charge to mass ratio is still lower.  Since the modulation potential is proportional to the charge to mass ratio, the modulation effect on heavier nuclei is even milder.  Considering that we use the ratio of their fluxes, most of the modulation effect cancels since they are positively charged.  So the modulation effect on heavier nuclei can also be safely absorbed in the systematic uncertainties.


To be able to compare with experimental data, we set the positron (electron) modulation potential in GalProp to $\phi^+ = 442$ (2) MV.  These values were determined by \citep{Gast:2009} for PAMELA.  \cite{2011JGRA..11602104U} showed that the time dependence of the solar modulation potential is not substantial over the period of PAMELA's data taking, and about the same average values can be used for Fermi-LAT.  We set the rest of the GalProp parameters to the values promoted by \cite{Strong:2004de}.  
%

\subsection{Statistical and numerical issues}


In order to extract the most favored values of the propagation parameters we have to calculate the posterior distribution $\mathcal{P}(p_i|D)$using suitable Bayesian priors $\mathcal{P}(p_j)$.  Assuming no prior knowledge justifies the use of uniform priors.  Since we have a previous knowledge about the order of magnitude of our parameters we use uniform priors for the propagation parameters (rather than for some functions, such as log, of them).  For the nuisance parameters prior knowledge is available in the form of a scan over GalProp predictions which are based on local measurements of cosmic ray fluxes different from those listed in Table \ref{tab:data}.  Thus for our nuisance parameters we use normally distributed priors. 


When evaluating $\sigma_i$ for the log likelihood in Eq.(\ref{eq:chi2}), following \cite{Trotta:2010mx}, we ignore theoretical uncertainties and combine statistical and systematic experimental uncertainties in quadrature
\begin{eqnarray}
 \label{eq:sigma1}
 \sigma_i^2 = \sigma_{i, statistical}^2 + \sigma_{i, systematic}^2.
\end{eqnarray}
This can be done for Fermi-LAT and the latest PAMELA $e^-$ flux.  Unfortunately, systematic uncertainties are not available for the rest of the cosmic ray measurements.  When this is the case, as an estimate of the systematics, we define $\sigma_i$ as the rescaled statistical uncertainty
\begin{eqnarray}
 \label{eq:sigma2}
 \sigma_i^2 = \sigma_{i, statistical}^2/\tau_i .
\end{eqnarray}
For simplicity, in this study, we use the same scale factor $\tau_i$ for all data points where systematic uncertainty is not available.  To remain mostly consistent with the work of \cite{Trotta:2010mx}, we set this common scale factor to a conservative value that they use: $\tau_i = 0.2$.  We checked that our conclusions only mildly depend on this choice.

We note that systematic errors in the data are not necessarily normally distributed point-to-point errors.  In fact, typically systematic errors are correlated, such as a systematic shift in the energy scale, and could be described by various probability distributions other than a Gaussian.  Unfortunately, these probability distributions are not provided by even those experimental collaborations that indicate a confidence interval for their systematic errors.  In the lack of this information, we use the simplest ansatz  which is adopted by most authors in the literature.  This estimate of the systematic errors is a simplified approximation of a more complicated situation. Nevertheless, for astrophysical data it captures the essence of systematic uncertainties.  After all, the simplest cosmic ray flux is a falling power law spectrum.  For this case a systematic shift in the energy scale, for example, can be re-interpreted as a systematic normalization shift of the spectrum.  Part of this shift is absorbed by our normalization nuisance parameters and part of it is approximated as Gaussian error.\footnote{We thank the referee of our manuscript to point out this issue.}
 


Due to the simplicity of the posterior density \citep{Trotta:2010mx} and its relatively low dimensionality we sample the parameter space $P$ and $P_{nuisance}$ according to a simple algorithm.  We select random model points from the parameter space according to a uniform distribution for $P$ and normally distributed for $P_{nuisance}$.  While this sampling technique is less efficient than the Monte Carlo based ones it enables us to trivially parallelize the numerical calculation.  It also allows us to simply check the robustness of our results against the change of certain assumptions such as the prior, the scale factor $\tau$ or the adequateness of the sampling. 


The simplicity of the likelihood function and the high number of data points used in this analysis also makes convergence testing relatively simple.  To test the validity of our results we can evaluate an approximate value of the posterior means, variances and the evidence adopting the procedure described by \cite{Tierney1986}.  To assure the adequacy of the sampling we can simply increase the number of samples of the posterior density until the numerically calculated evidence is within 5 \% of the one obtained by the Laplace method.
During this procedure we found that to extract the posterior probabilities presented in this paper about one million samples of the posterior density were required over the parameter space in Eq.(\ref{eq:P}) and (\ref{eq:Pn}).  The gathering of this sample consumed about $2 \times 10^5$ CPU hours.

\subsection{Experimental data}
\label{sec:ExpDat}

We included 219 of the most recent experimental data points in our statistical analysis.  These contained 114 electron-positron related, and 105 Boron/Carbon, anti-proton/proton, (Sc+Ti+V)/Fe and Be-10/Be-9 cosmic ray flux measurements.  As a number of experiments have energy ranges which overlap, we chose the most recent experimental data points in those energy ranges.  

\begin{table}
\begin{center}
\caption{Cosmic ray experiments and their energy ranges over which we have chosen the data points for our analysis.  We split the data into two groups: electron-positron flux related (first five lines in the table), and the rest.  We perform two independent Bayesian analyses to show the significant tension between the two data sets.
\vspace{5mm}
\label{tab:data}}
\begin{tabular}{lllc}
\hline
\hline
\bf Measured flux  & \bf Experiment                                & \bf Energy   & \bf Number of   \\
                   &                                               & (GeV)        & \bf data points \\
\hline
\hline
                   & AMS \citep{Aguilar:2002ad}                     & 0.60 - 0.91  & 3  \\
 $e^+ + e^-$       & Fermi-LAT \citep{Ackermann:2010ij}             & 7.05 - 886   & 47 \\
                   & HESS \citep{Aharonian:2008aa,Aharonian:2009ah} & 918 - 3480   & 9  \\
\hline                                                                             
 $e^+/(e^+ + e^-)$ & PAMELA \citep{Adriani:2010ib}                  & 1.65 - 82.40 & 16 \\
\hline    
 $e^-$             & PAMELA \citep{Adriani:2011xv}                  & 1.11 - 491.4 & 39 \\
\hline
\hline                                                                             
 anti-proton/proton & PAMELA \citep{Adriani:2010rc}                  & 0.28 - 129   & 23 \\
\hline                                                                            
                   & IMP8 \citep{Moskalenko:2001ya}                 & 0.03 - 0.11  &  7 \\
                   & ISEE3 \citep{1988ApJ...328..940K}              & 0.12 - 0.18  &  6 \\
 Boron/Carbon      & \cite{1978ApJ...223..676L} & 0.30 - 0.50  &  2 \\
                   & HEAO3 \citep{1990AA...233...96E}               & 0.62 - 0.99  &  3 \\
                   & PAMELA \citep{PAMELABORONCARBON}               & 1.24 - 72.36 &  8 \\
                   & CREAM \citep{Ahn:2008my}                       &   91 - 1433  &  3 \\
\hline                                                                             
 (Sc+Ti+V)/Fe      & ACE \citep{2000AIPC..528..421D}                & 0.14 - 35    & 20 \\
                   & SANRIKU \citep{1999ICRC....3..105H}            & 46 - 460     &  6 \\
\hline                                                                             
                   & \cite{1980ApJ...239L.139W} & 0.003 - 0.029 &  3 \\
                   & \cite{1981ICRC....9..195G}  & 0.034 - 0.034 &  1 \\
                   & \cite{1980ApJ...239L.139W}       & 0.06 - 0.06   &  1 \\
 Be-10/Be-9        & ISOMAX98 \cite{2001ICRC....5.1655H}             & 0.08 - 0.08   &  1 \\
                   & ACE-CRIS \citep{davis:421}                       & 0.11 - 0.11   &  1 \\
                   & ACE \citep{2001AdSpR..27..727Y}                  & 0.13 - 0.13   &  1 \\
                   & AMS-02 \citep{2004EPJC...33S.941B2}              & 0.15 - 9.03   & 15 \\
\hline                                                                         
\hline
\end{tabular}
\end{center}
\end{table}

For $e^+ + e^-$ we used the most recent data from AMS by \cite{Aguilar:2002ad}, Fermi-LAT by \cite{Ackermann:2010ij} and HESS by \cite{Aharonian:2008aa, Aharonian:2009ah}.  The energy ranges in which we use each experiment is listed in Table \ref{tab:data}.  The AMS experiment reported an excess in high energy positrons for energies greater than 10 GeV.  The Fermi-LAT collaboration reported a high precision measurement of the $e^+ + e^-$ spectrum for energies from 7 GeV to 1 TeV using its Large Area Telescope (LAT).  This spectrum extended their previously published electron positron spectrum over an energy range of 20 GeV to 1 TeV \citep{Abdo:2009zk} and is flatter than results reported by earlier experiments.  HESS's atmospheric Cherenkov telescope (ACT) reported a significant steepening of the electron plus photon spectrum above one TeV.

The PAMELA collaboration measured the flux of the positron fraction $e^+/(e^+ + e^-)$, between 1.5 and 100 GeV \citep{Adriani:2008zr}.  They observed that this differential positron fraction falls slower than expected for energies above 10 GeV.  This behaviour is different from that of the background of secondary positrons produced during propagation of cosmic rays in the Galaxy.  Recently PAMELA released the measurement of the $e^-$ flux alone \citep{Adriani:2011xv} robustly confirming the $e^+ + e^-$ spectrum by Fermi-LAT.

Cosmic ray anti-protons can be used to study the production of primary and secondary cosmic rays and their transport throughout the Galaxy.  Detailed anti-proton spectra requires a large number of measurements over a larger energy range, with good statistics.  Previous balloon borne experiments such as CAPRISE98 \citep{Boezio:2001ac} and HEAT \citep{2001PhRvL..87A1101B} detected only a small number of anti-protons with limited statistics.  The PAMELA satellite experiment \citep{Adriani:2010rc} provided a comprehensive measurement of the anti-proton/proton flux ratio for an energy range of 1-100 GeV.  PAMELA's spectrum follows the same trend as other recent anti-proton/proton ratio measurements.  The energy range over which we use the PAMELA experiment for the anti-proton/proton ratio is listed in Table \ref{tab:data}.

In comparison to primary/primary or secondary/secondary cosmic ray ratios, stable secondary to primary cosmic ray ratios, such as Boron/Carbon and (Sc+Ti+V)/Fe ratio, are the most sensitive to variation in the propagation of cosmic rays in the Galaxy. Their sensitivity arises from the fact that primary cosmic rays are generated by the original source while secondary cosmic rays are created by the interaction of their primaries with the interstellar medium \citep{2008ICRC....2..183C}. Primary/primary and secondary/secondary cosmic ray ratios have a low sensitivity to variation in the propagation parameters as the denominator and numerator are produced by similar propagation mechanisms. The Boron/Carbon and (Sc+Ti+V)/Fe ratio provide an indication (over different energy ranges) of the amount of interstellar material that primary cosmic rays traverse as a function of energy \citep{2008ICRC....2..183C}. The experiments used to define the B/C and (Sc+Ti+V)/Fe ratio for our analysis are found in Table \ref{tab:data}.


In conjunction with stable secondary/primary ratios such as the Boron/Carbon ratio, unstable isotope ratios such as Beryllium-10/Beryllium-9 can be used to constrain the time it takes for cosmic rays to propagate through the Galaxy \citep{Malinin:2004pw}.  In this work we use Be-10/Be-9 data from various experiments, such as ISOMAX98 \citep{2001ICRC....5.1655H}, ACE-CRIS \citep{davis:421}, ACE \citep{2001AdSpR..27..727Y} and AMS-02 \citep{2004EPJC...33S.941B2}.

\section{Results}

\subsection{Is there a cosmic ray anomaly?}


We begin our results section by investigating the question whether the present cosmic ray data can be used to justify the existence of an anomaly in the cosmic electron-positron spectrum. Both the reality of an anomaly in the PAMELA $e^+/(e^+ + e^-)$ flux and the absence of such in the anti-proton flux have been questioned by \cite{Katz:2009yd} and \cite{Kane:2009if}, respectively.  Recently \cite{Trotta:2010mx} argued that the Fermi-LAT data can be well matched by the diffusion model, as encoded in GalProp, simply by adjusting the parameters of the propagation model.  Their Fig. 8 clearly shows that the Fermi-LAT data agree reasonably well with the propagation model that was their best fit to 76 cosmic ray spectral data points.  \cite{Trotta:2010mx} also acknowledge that the ``positron fraction, shown in the right panel of Fig. 8, does not agree with the PAMELA data (Adriani et al. 2009), but this was expected since secondary positron production in the general ISM is not capable of producing an abundance that rises with energy''.  In other words, they conclude that PAMELA cannot be fitted by simply adjusting the propagation parameters.  We take this as an important indication that the cosmic ray anomaly is real and requires a detailed investigation rather than the adjustment of the propagation model to explain it. 


\begin{figure}
\begin{center}
\includegraphics[width=0.47\textwidth]{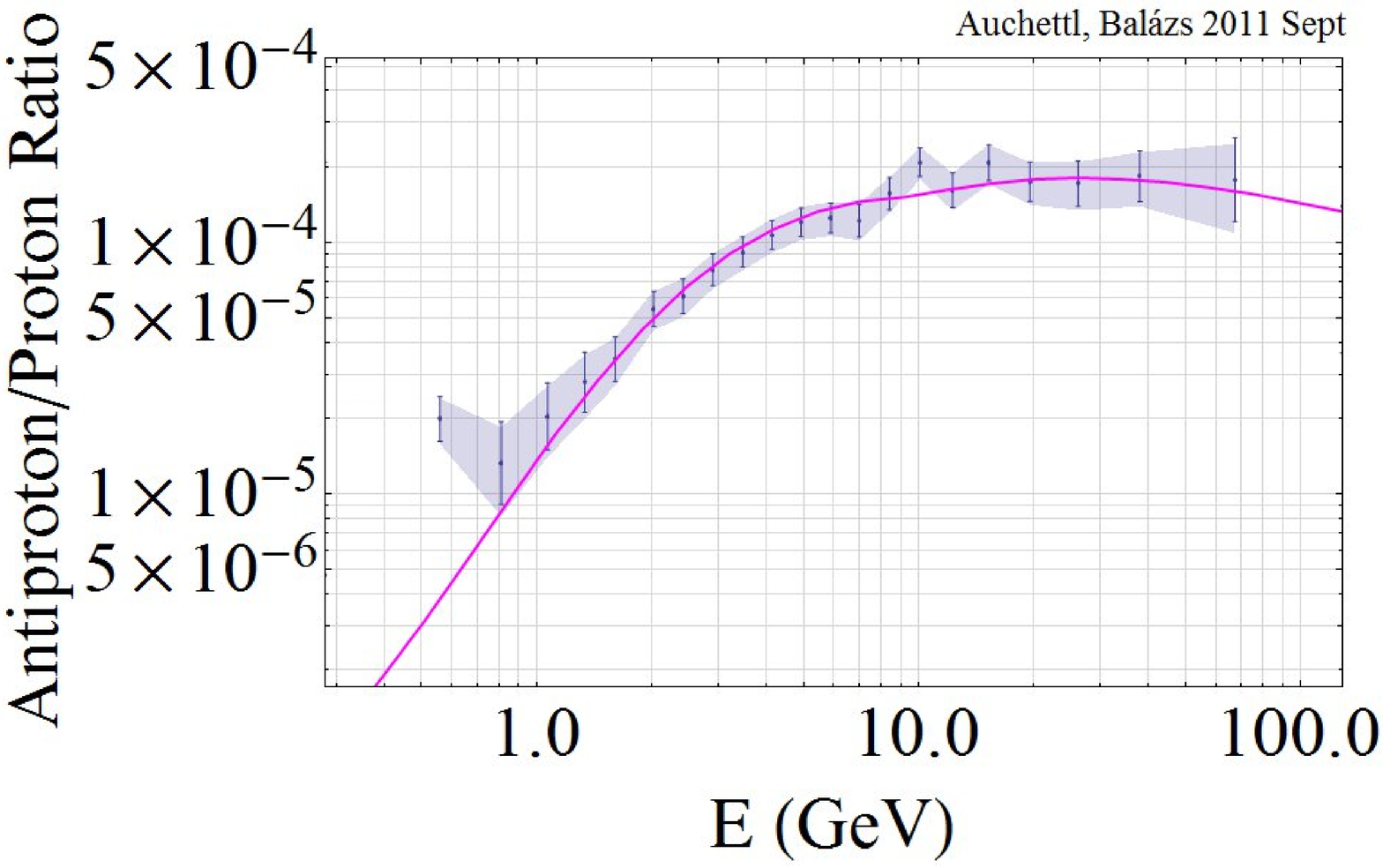}\vspace{3mm}
\includegraphics[width=0.47\textwidth]{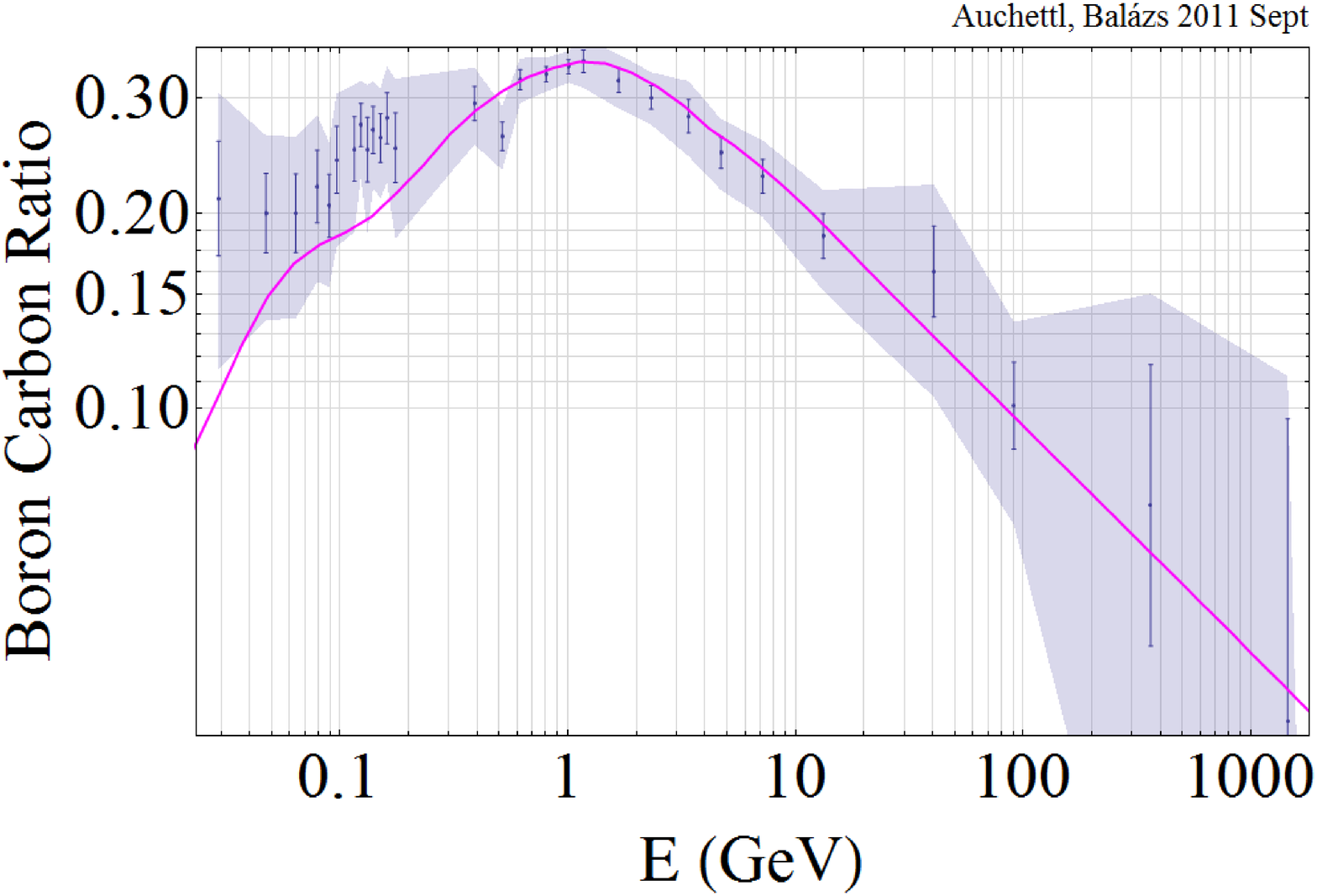}
\includegraphics[width=0.47\textwidth]{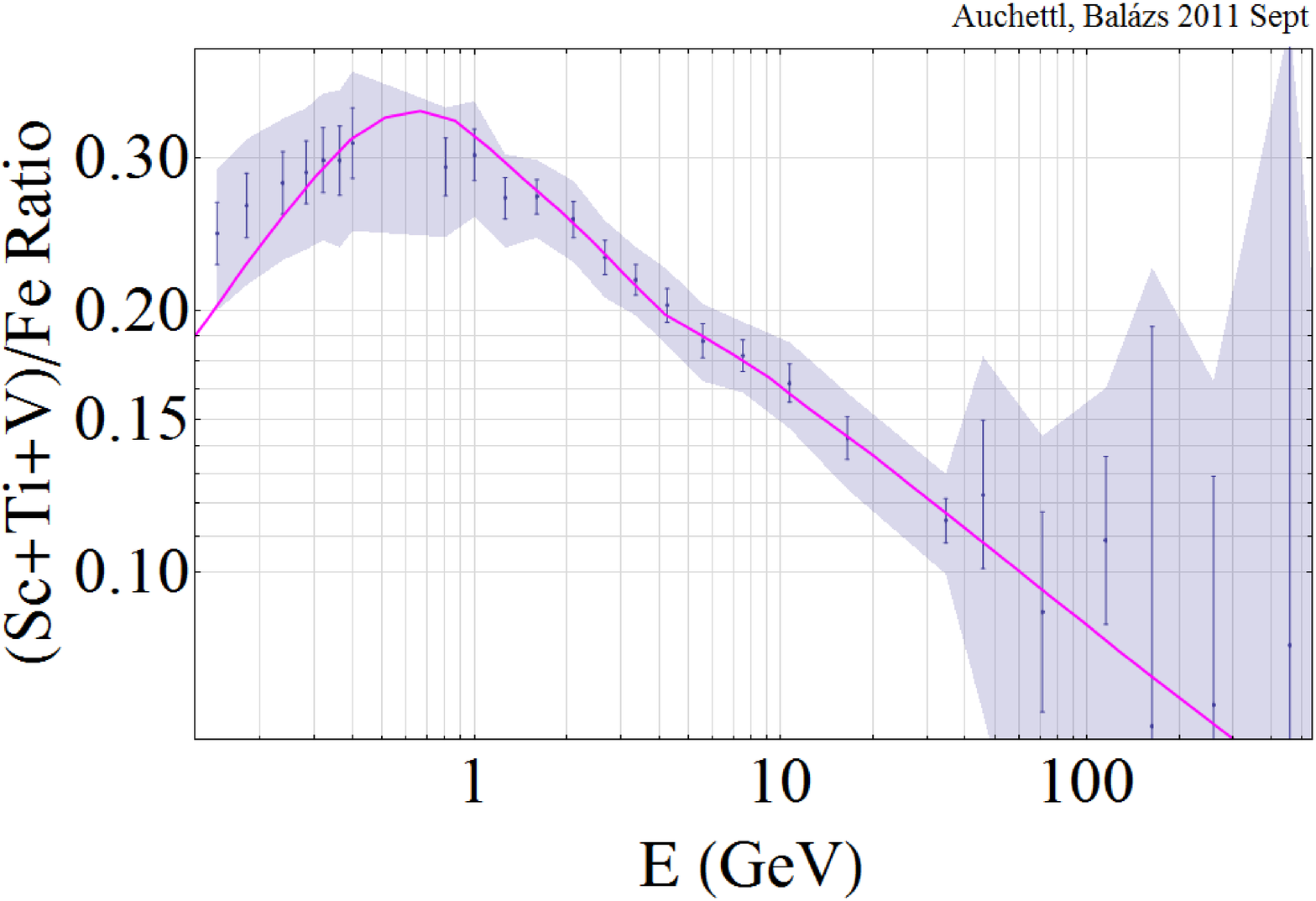}
\includegraphics[width=0.47\textwidth]{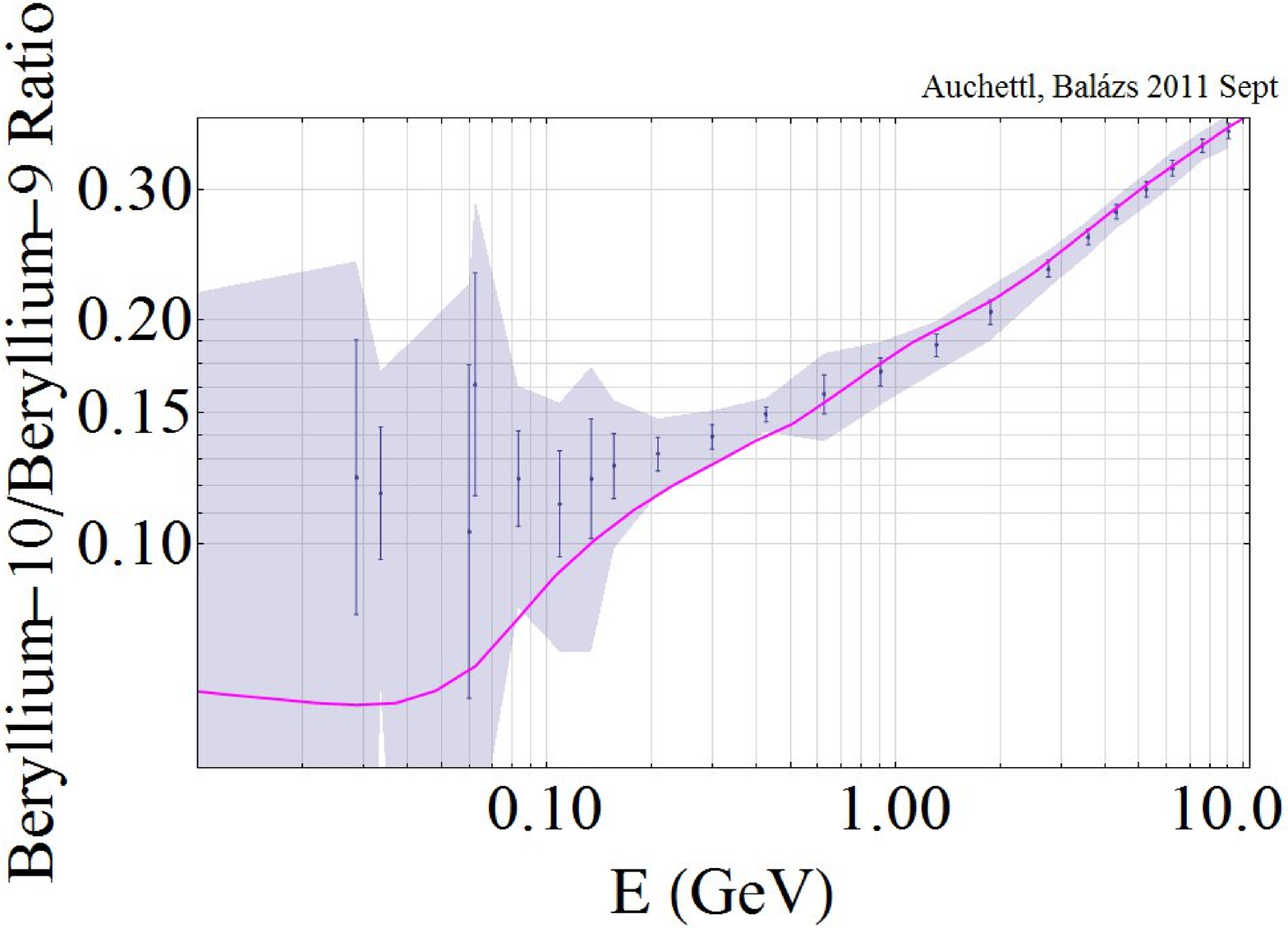}
\end{center}
\caption{Best fit curves compared to non-electron-positron related data.  The curves were calculated using the most probable parameter values inferred from the ${\bar {\rm p}}$/p, B/C, (Sc+Ti+V)/Fe and Be data.  These most probable values correspond to the peak values of the posterior probabilities shown in red in Fig. \ref{fig:Posteriors}.  The best fit curves pass through the estimated systematic error bands, shown in gray.}
\label{fig:BestFits}
\end{figure}

The hypothesis that the adjustment of the propagation parameters does not solve the cosmic ray anomaly is further supported by the fact that not all cosmic ray data can be fitted well with a single set of these parameters.  It is already evident from Fig. 7 and 8 of \cite{Trotta:2010mx} that the best-fit of the propagation parameters to the rest of the cosmic ray data does not fit well AMS, Fermi and HESS simultaneously.  This is exactly what we find in our analysis.  Our best fit for all cosmic ray data excluding AMS, Fermi, HESS and PAMELA data with electron and/or positron fluxes gives a $\chi^2$ per degree of freedom of 0.34.  As a consequence the best fit curves all pass through the estimated systematic error bands, shown in gray, in Fig. \ref{fig:BestFits}.  When this fit is compared to the AMS, Fermi, HESS and PAMELA electron and/or positron flux the $\chi^2$ per degree of freedom we obtain is 24, which signals considerable tension bordering exclusion.  The converse also holds.  By changing the propagation parameters, we can find an ideal fit for the electron-positron related fluxes with $\chi^2$ per degree of freedom of 1.0.  But for the rest of the cosmic ray data the same fit results in a $\chi^2$ per degree of freedom of 3.1, which is a significant pull for 105 degrees of freedom.  These discrepancies signal a statistically significant tension between the electron-positron measurements and the rest of the comic ray data.  


\begin{figure}
\begin{center}
\includegraphics[width=0.47\textwidth]{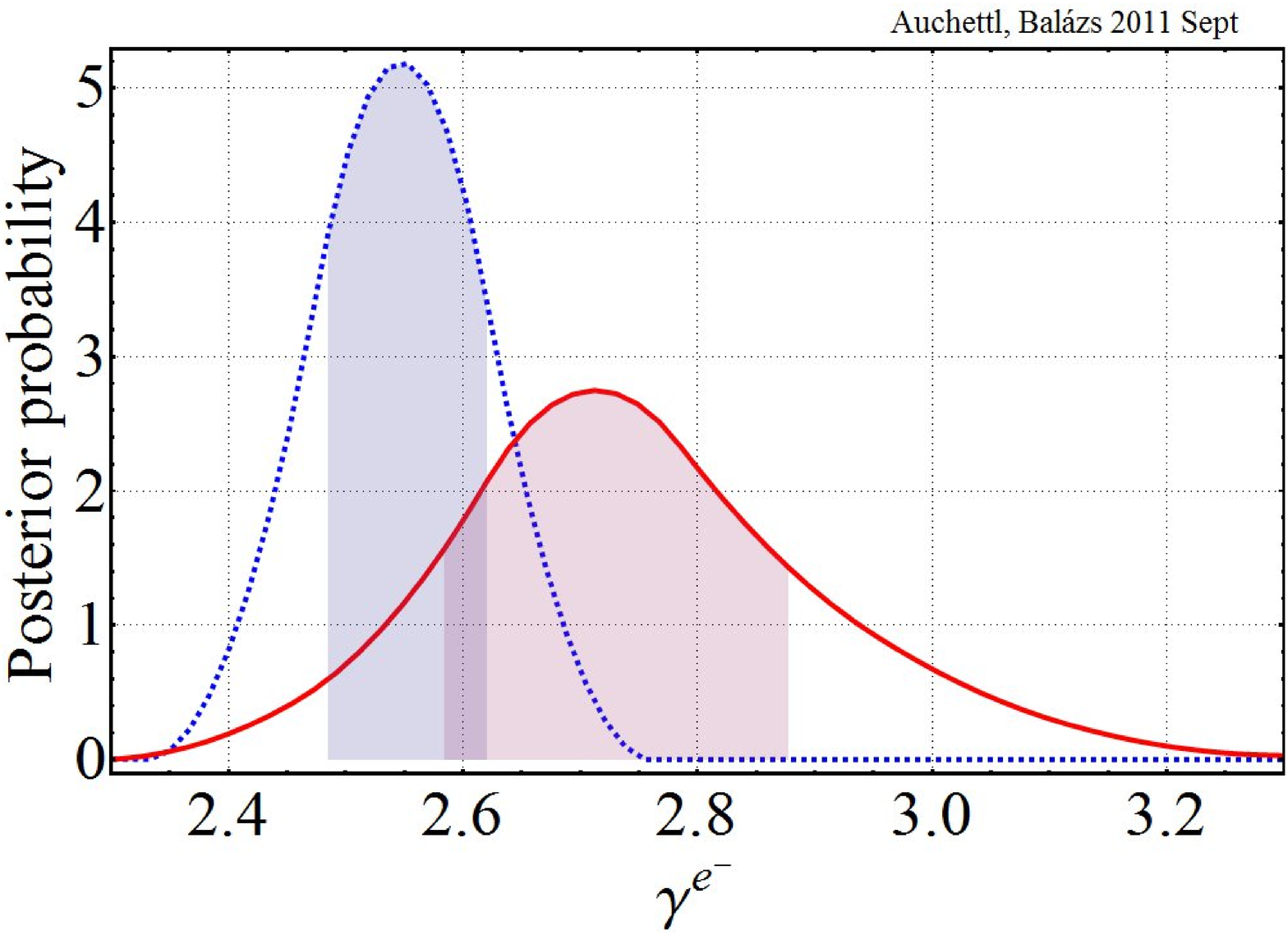}\vspace{3mm}
\includegraphics[width=0.47\textwidth]{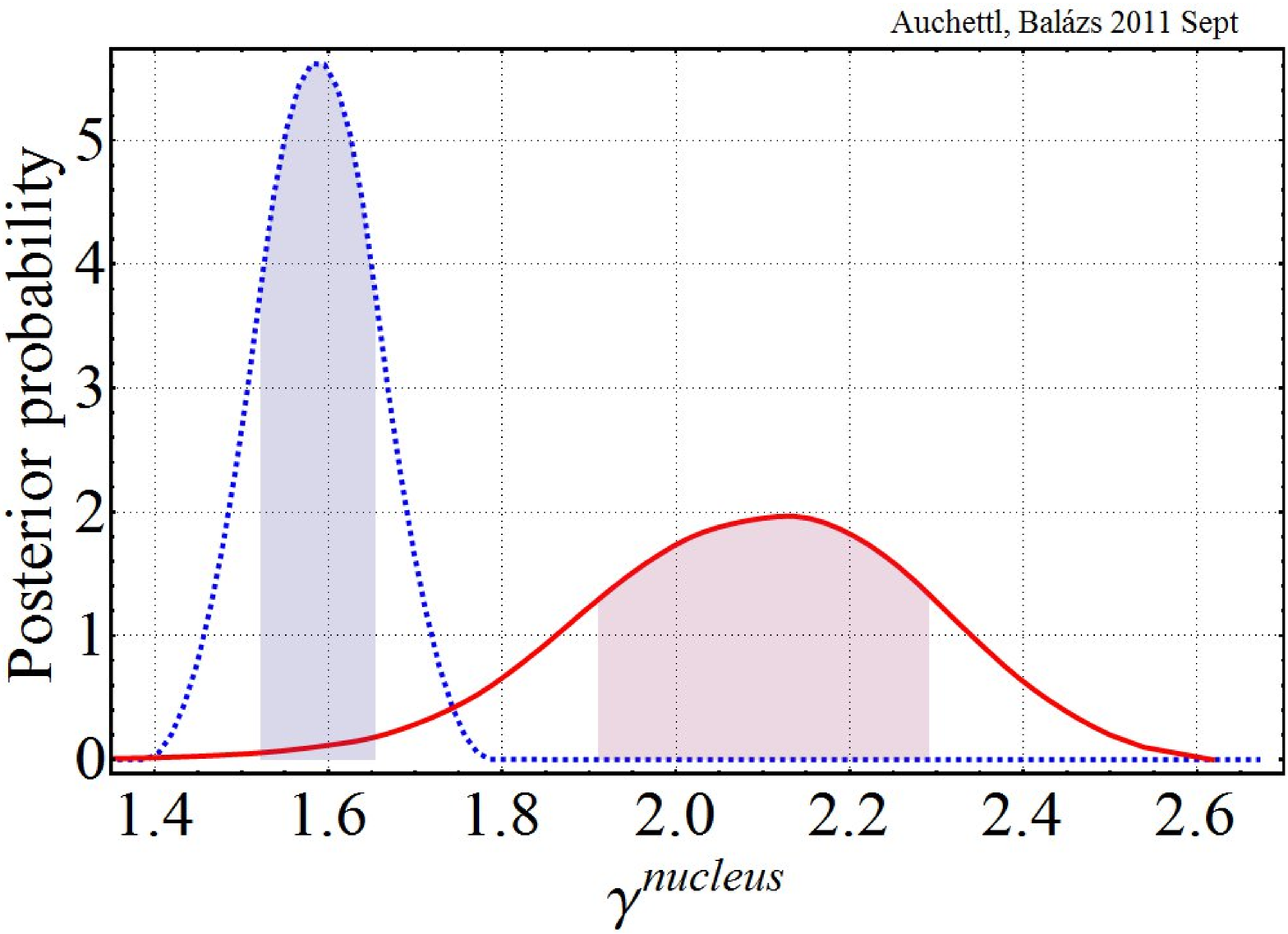}
\includegraphics[width=0.47\textwidth]{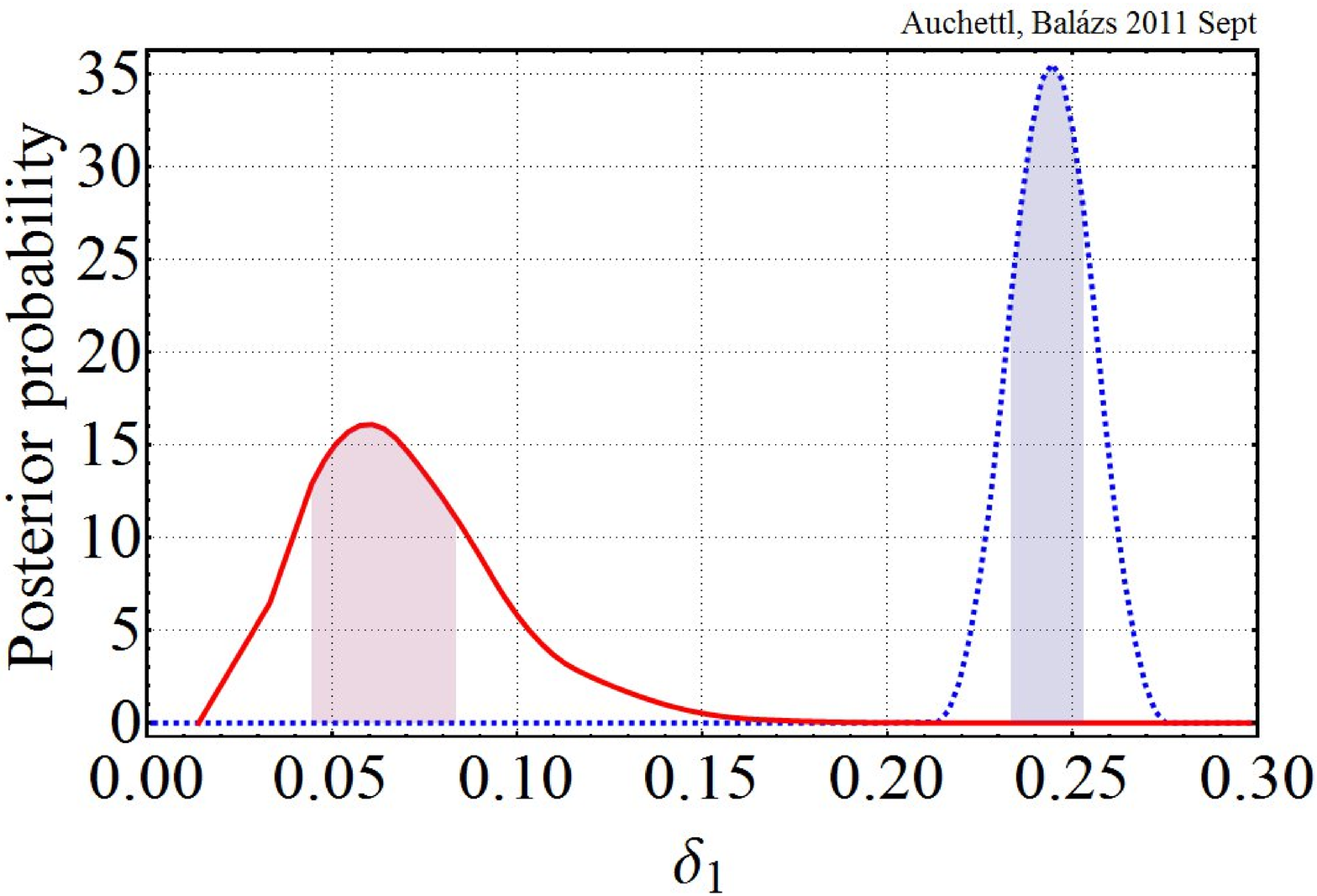}\vspace{3mm}
\includegraphics[width=0.47\textwidth]{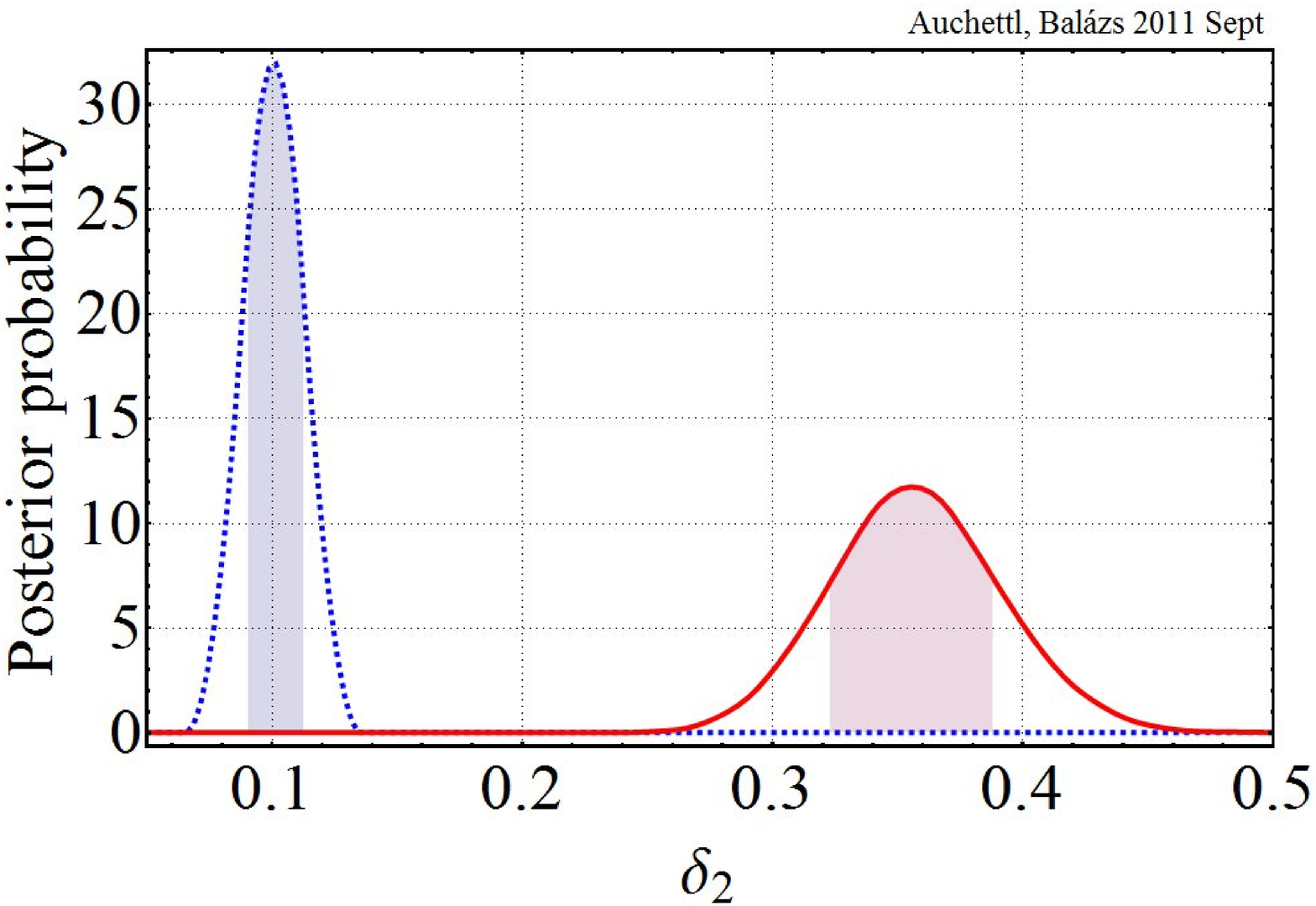}
\includegraphics[width=0.47\textwidth]{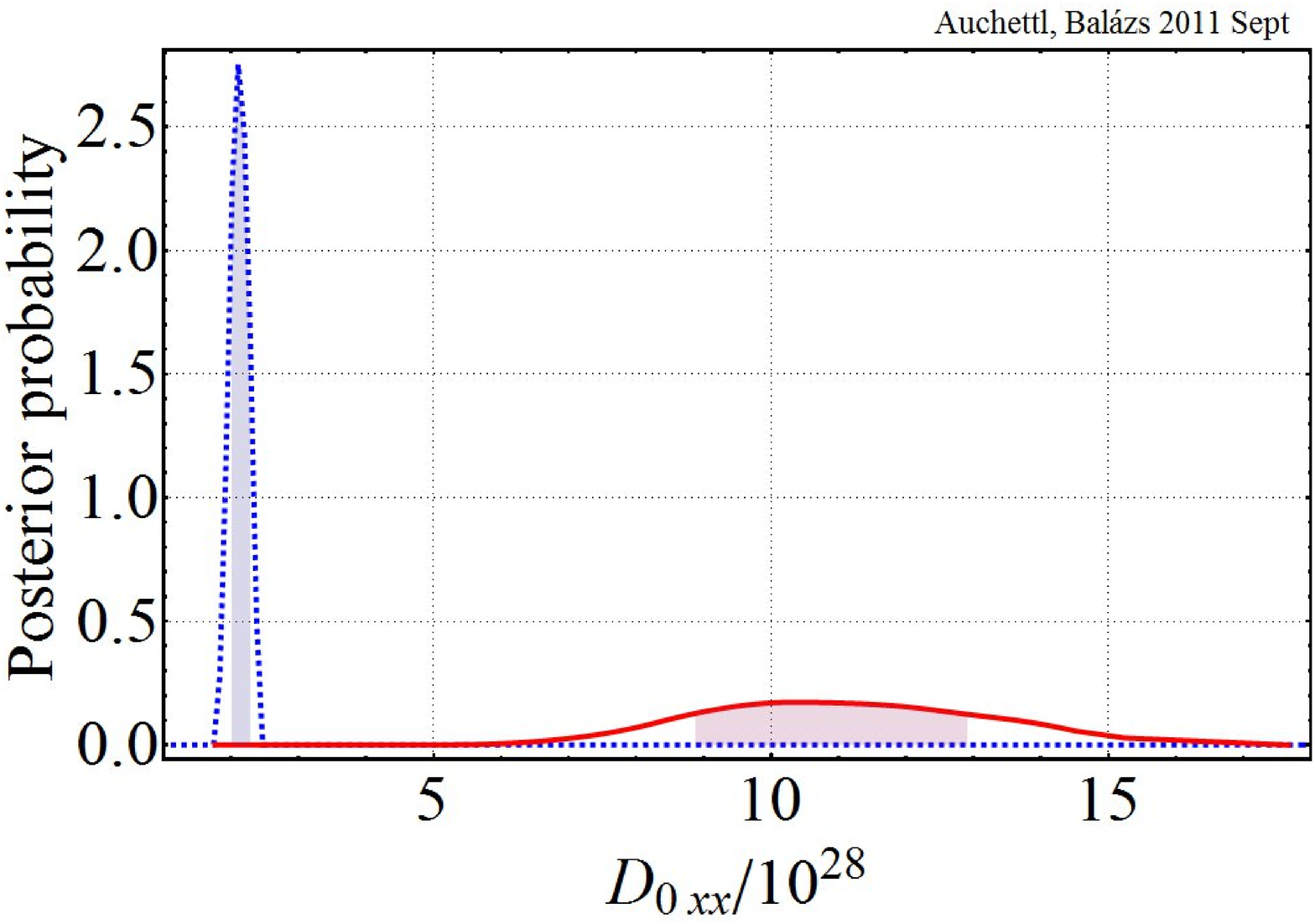}
\end{center}
\caption{Marginalized posterior probability distributions of propagation parameters listed in Eq.(\ref{eq:P}).  The dashed blue curves show results with likelihood functions containing electron and/or positron flux data while the likelihood functions for the solid red curves contain only the rest of the comic ray data.  Shaded areas show the 68 \% credibility regions.  A statistically significant tension between the electron-positron data and the rest is evident in the three lower frames.}
\label{fig:Posteriors}
\end{figure}

To further investigate the tension, we divide the cosmic ray data into two groups: 114 measurements containing observations of electron and/or positron fluxes (AMS, Fermi, HESS, and PAMELA) and the rest of 105 data points (anti-proton/proton, Boron/Carbon, (Sc+Ti+V)/Fe, Be-10/Be-9).  We perform a Bayesian analysis independently on these two sets of data extracting the preferred values of the propagation parameters.  Remarkably, we found that we can obtain information about the electron-positron related propagation parameters from the rest of the data because the propagation of the cosmic rays is entangled for several reasons.  First, certain propagation parameters, most importantly for us $D_{0xx}$, are species independent.  Second, the transport equation includes nuclear fragmentation and decay, which directly contributes to the secondary electron-positron fluxes.  Third, since their energy density is comparable to the interstellar radiation and magnetic fields, various species of cosmic rays affect each other dynamics.  

Due to the correlations pointed out above certain parameters of the electron-positron propagation are constrained even if no electron-positron related data is used in our fit.  Unfortunately, the injection indices remain virtually unconstrained.  In order to fix those parameters we resorted to use a minimal amount of information from the electron-positron related fluxes.  We decided to use data points from the $e^- + e^+$ spectrum because (1) they span the widest energy range and the end points of the $e^- + e^+$ spectrum, partially due to their high uncertainty, appeared to agree with the theoretical predictions even before we set out to find the most optimal parameter values; (2) in the low energy region they are relatively insensitive for solar modulation effects; and (3) because in the mid energy range the $e^- + e^+$ theoretical prediction develops an insensitivity to the values of the propagation parameters (c.f. the distinct bow-tie shape of the theory uncertainty band).

With this in mind, we included four $e^\pm$ related data points in the analysis together with the ${\bar {\rm p}}$/p, B/C, (Sc+Ti+V)/Fe and Be data.  These were the lowest energy point of AMS, the highest energy point of HESS, and the 19.40 GeV and 29.20 GeV data points of Fermi-LAT.  We have checked that our result are robust against this choice and do not bias the final conclusion.


Fig. \ref{fig:Posteriors} clearly shows that the two subsets of cosmic ray data are inconsistent with the hypothesis that the cosmic ray propagation model and/or sources implemented in GalProp provides a good theoretical description.  The five frames display the marginalized posterior probability densities of our selected propagation parameters.  Dashed blue curves show results with likelihood functions containing only electron-positron related flux data (AMS, Fermi, HESS, and PAMELA) while the likelihood functions for the solid red curves contain only the rest of the comic rays (anti-proton/proton, Boron/Carbon, (Sc+Ti+V)/Fe, Be-10/Be-9).  Shaded areas show the 68 \% credibility regions for the parameters.  Table \ref{tab:bestfit} shows the numerical values of the best fits and the 68 \% credibility ranges for the propagation parameters.


\begin{table}[ht]
\begin{center}
\caption{Best fit values of the propagation parameters and their 68 \% credibility ranges.  Numerical values are shown for both fits: including the electron-positron related cosmic ray data only, and including the rest of the data.
\vspace{5mm}
\label{tab:bestfit}}
\begin{tabular}{lcccccc}
\hline
\hline
 parameter & \multicolumn{2}{c}{Fit for the $e^\pm$ related data} 
           & \multicolumn{2}{c}{Fit for the rest of the data} \\
           &  best fit value & 68\% Cr range &  best fit value & 68\% Cr range \\
\hline
$\gamma^{e^-}$               & 2.55 & \{2.45, 2.60\} &  2.71 & \{2.54, 2.92\} \\
$\gamma^{nucleus}$           & 1.60 & \{1.51, 1.69\} &  2.10 & \{1.88, 2.92\} \\
$\delta_1$                   & 0.24 & \{0.23, 0.26\} &  0.06 & \{0.04, 0.08\} \\
$\delta_2$                   & 0.10 & \{0.08, 0.12\} &  0.35 & \{0.32, 0.39\} \\
$D_{0xx}$ [$\times 10^{28}$] & 2.17 & \{1.85, 2.19\} & 11.49 & \{8.86,13.48\} \\
\hline
\hline
\end{tabular}
\end{center}
\end{table}

In the first two frames, showing the posterior densities of the electron and nucleus injection indices $\gamma^{e^-}$ and $\gamma^{nucleus}$, there is a mild but tolerable tension between the electron-positron related and the rest of the cosmic ray data.  The last three frames, on the other hand, indicate statistically significant tension between the $e^+$-$e^-$ and the rest of the data.  The 68 \% credibility regions for the spatial diffusion coefficients $\delta_1$ and $\delta_2$ and that of $D_{0xx}$ fall far away from each other when determined using the two different cosmic ray data sets.  Although it is not shown, it is easily inferred that not even the 99 \% credibility regions overlap.  It appears that by adjusting the cosmic ray parameters we can, indeed, achieve a good fit to either the electron-positron related fluxes or to the rest of the data but not to both simultaneously.

Our interpretation of the tension between the electron-positron fluxes and the rest of the cosmic ray data is the following.  The measurements of PAMELA and Fermi-LAT are affected by new physics which is unaccounted for by the propagation model and/or cosmic ray sources included in our calculation.  We base this hypothesis partly on the earlier quoted statement of \cite{Trotta:2010mx} that ``secondary positron production in the general ISM is not capable of producing an abundance that rises with energy''.  The behavior of the PAMELA $e^+/(e^+ + e^-)$ data is unexpected based on general theoretical principles and when it is fit by adjusting the propagation parameters it leads to a bad fit to the rest of the data.  An anomaly in PAMELA $e^+/(e^+ + e^-)$ is also expected to produce an anomaly in the Fermi-LAT $e^+ + e^-$ and the PAMELA $e^-$ spectra.

We note that the recently released PAMELA $e^-$ flux \citep{Adriani:2011xv}, which is included among our electron-positron related data, considerably increases this tension.  We checked that without the inclusion of the PAMELA $e^-$ flux the tension is noticeably milder.  This, and the effect of the extra data that we use, probably explains why this tension was not detected by \cite{Trotta:2010mx}.

\subsection{What is the size of the anomaly?}

We attempt to extract the size of the new physics signal, after arriving to the conclusion that new physics is buried in the electron-positron fluxes.  Based on our findings our working hypothesis is that the new physics is affecting the electron-positron fluxes but hardly influences the rest of the cosmic rays.  Under this hypothesis the cosmic ray propagation parameters can be determined from the unbiased data: anti-proton/proton, Boron/Carbon, (Sc+Ti+V)/Fe, Be-10/Be-9.  This means that we can use the central values and credibility regions of the parameters determined using this data to calculate a background prediction for all cosmic ray data including the electron-positron fluxes.  Once we quantify the background we can subtract it from the electron-positron data to see whether there is a statistically significant signal can be extracted.  

In the first step, we use the central values of the propagation parameters determined earlier using ${\bar {\rm p}}$/p, B/C, (Sc+Ti+V)/Fe, Be-10/Be-9 to calculate a central value prediction for the PAMELA and Fermi-LAT electron-positron fluxes.  Then we use all the scanned points in the parameter space lying within the 68 \% credibility region of all the five scanned parameters to establish a 1-$\sigma$ band around this central value.  We will refer to this band as the 1-$\sigma$ uncertainty of the background.  We overlay this uncertainty band on the Fermi-LAT electron+positron and the PAMELA electron and positron fraction fluxes.


\begin{figure}
\begin{center}
\includegraphics[width=0.47\textwidth]{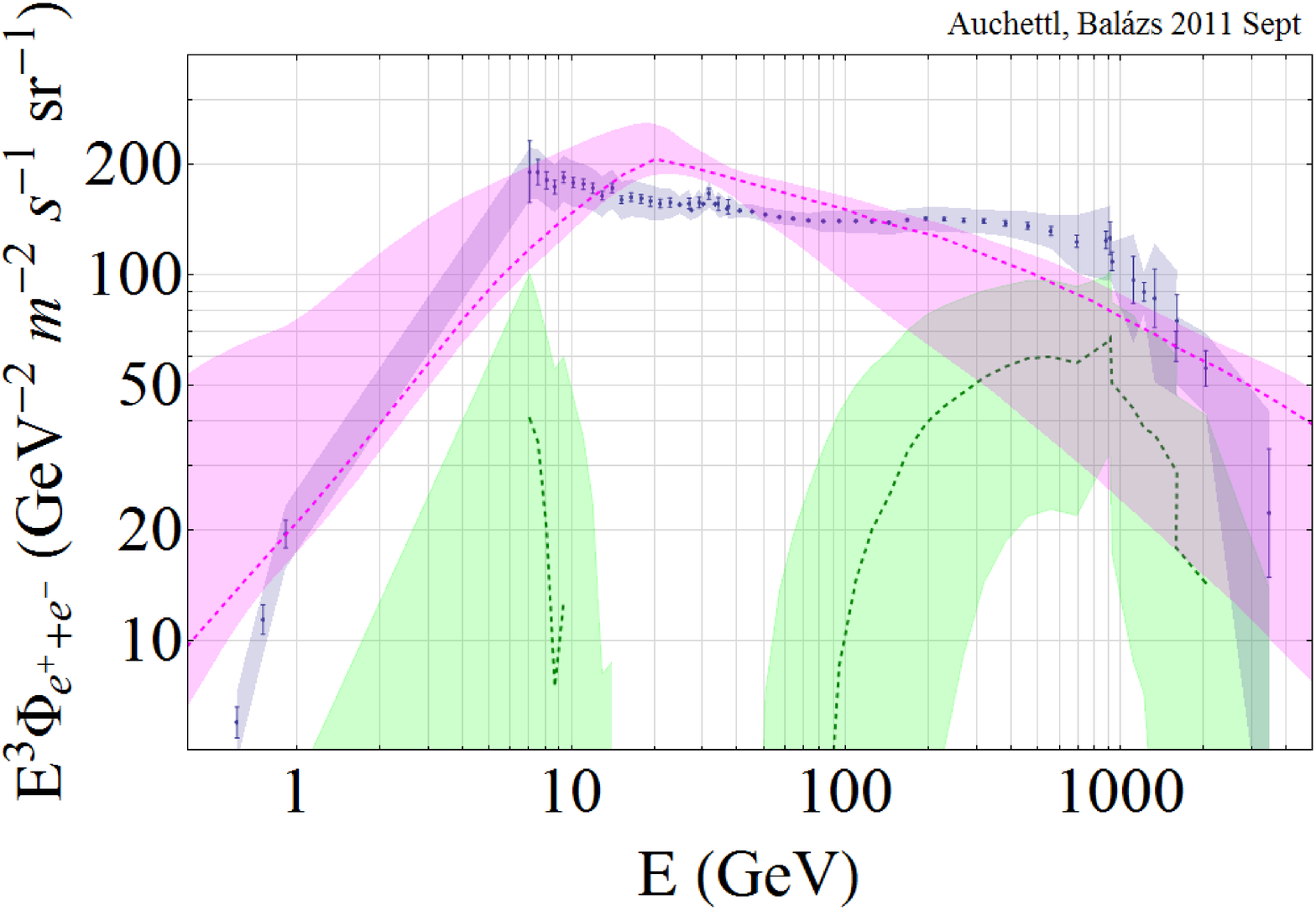}\vspace{3mm}
\includegraphics[width=0.47\textwidth]{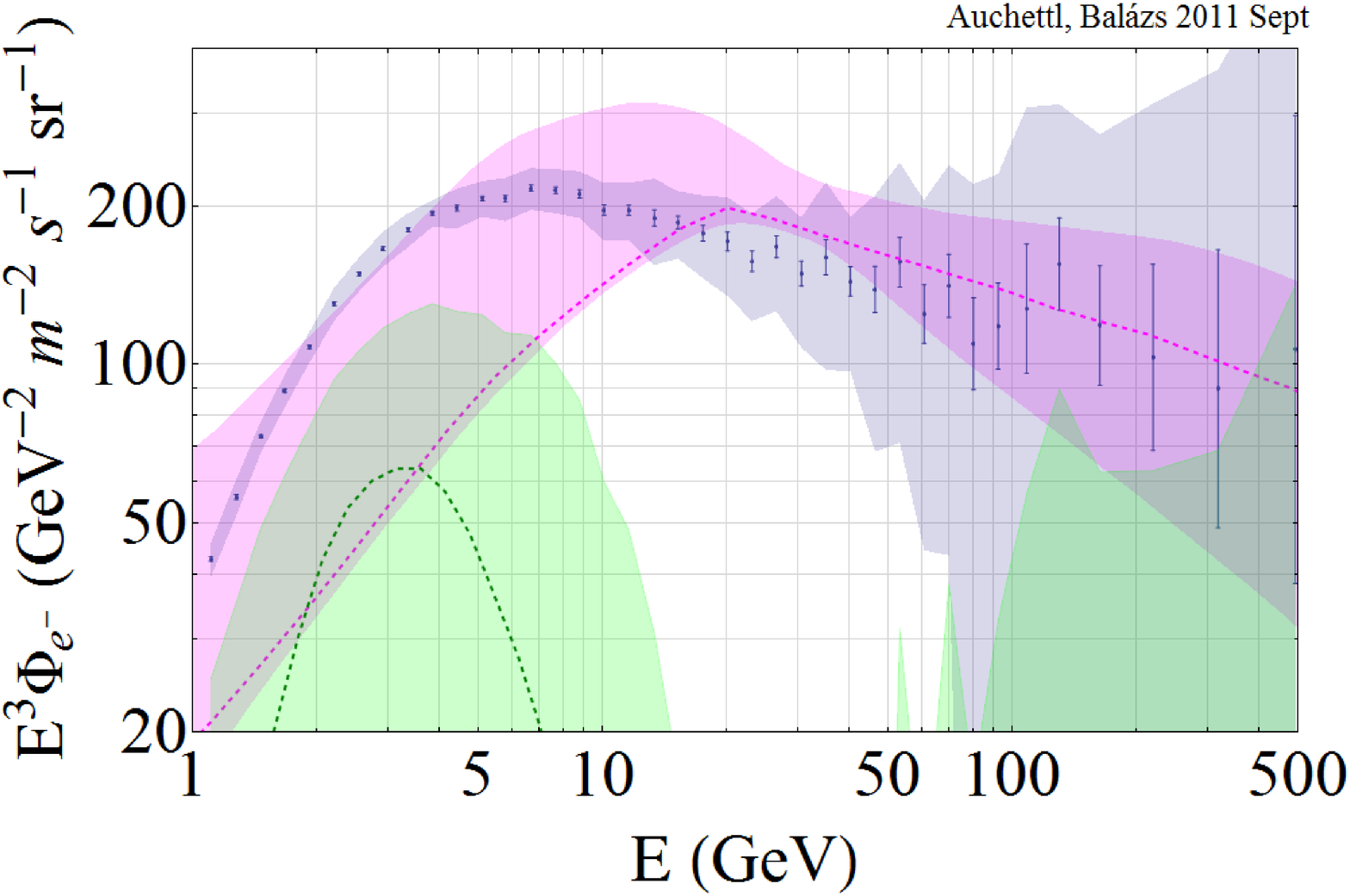}
\includegraphics[width=0.47\textwidth]{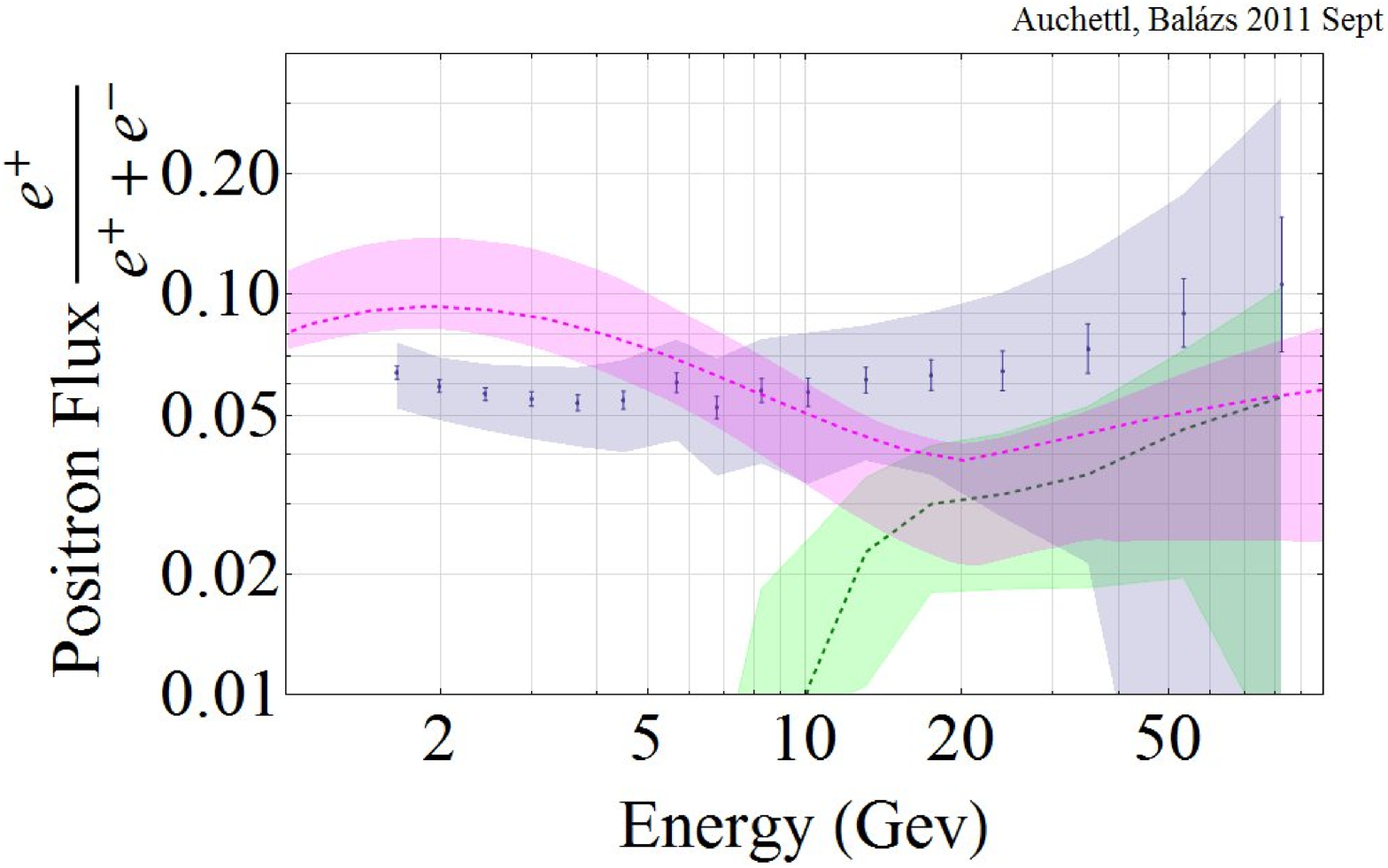}
\end{center}
\caption{Electron-positron fluxes measured by Fermi-LAT and PAMELA (gray bands) with the extracted size of the electron-positron anomaly (green bands).  Combined statistical and systematic uncertainties are shown for Fermi-LAT and PAMELA $e^-$, while ($\tau = 0.2$) scaled statistical uncertainties are shown for PAMELA $e^+/(e^++e^-)$.  Our background predictions (magenta bands) are also overlaid.}
\label{fig:Backgrounds}
\end{figure}

Fig. \ref{fig:Backgrounds} shows the comparison of the measured electron-positron fluxes and their backgrounds.  Statistical and systematic uncertainties combined in quadrature are shown for Fermi-LAT and PAMELA $e^-$, while ($\tau = 0.2$) scaled statistical uncertainties are shown for PAMELA $e^+/(e^++e^-)$ as gray bands.  Our background prediction is overlaid as magenta bands.  The central value and the 1-$\sigma$ uncertainty of the calculated anomaly is displayed as green dashed lines and bands.  As the first frame shows the Fermi-LAT measurements deviate from the predicted background both below 10 GeV and above 100 GeV.


As we shall discuss later, the low energy deviation might be due to the inadequacies of the propagation model, so here we concentrate on the deviation between the background and the measurements above 100 GeV.  In our interpretation this is a weak but statistically significant signal of the presence of new physics in the electron+positron flux.  Based on the difference between the central values of the data and the background a similar conclusion can be drawn from PAMELA.  Unfortunately, the sizable uncertainties for the PAMELA measurements prevent us to claim a statistically significant deviation.\footnote{Recently the Fermi collaboration revealed a very preliminary positron fraction measurement nicely confirming the PAMELA results \citep{FermiPosFrac}.  Even though the Fermi-LAT makes use of only the Earth's magnetic field, it appears to have less systematic uncertainties than that of PAMELA.  If the officially published Fermi-LAT measurement will indeed reduce the systematic errors to the level of PAMELA's statistical ones, our background will deviate from it revealing a signal also in the positron fraction.}

After having determined the background for the electron-positron fluxes, we can subtract the background from the measured flux to obtain the size of the new physics signal.  We determine the central value of the signal by subtracting the central value of the background from the central value of the data.  The 1-$\sigma$ uncertainty of the signal is the quadratically combined experimental and background uncertainty.  Results for the electron-positron anomaly are also shown in Fig. \ref{fig:Backgrounds}.  As expected based on the background predictions a non-vanishing anomaly can be established for the Fermi-LAT $e^+ + e^-$ flux, while no anomaly with statistical significance can be claimed for PAMELA due to the large uncertainties.

\subsection{What is the source of the anomaly?}

Based on the available evidence we can only speculate about the origin of the discrepancy between the data and predictions of the cosmic electron-positron spectra.  Since the publication of the first PAMELA positron fraction measurement by \cite{Adriani:2008zr} speculation has been abundant.  The first obvious assumption is that some aspect of the propagation model used in the present calculation is insufficient for the proper description of the electron-positron fluxes arriving at Earth \citep{Stawarz:2009ig, Donato:2010vm, Arakida:2010vz, Tawfik:2010qf}.  In this case there exists no anomaly in the data.  One such plausible effect, which is missed by the 2-dimensional calculation in GalProp, is the spectral hardening of cosmic rays caused by non-steady sources \citep{1979ApJ...228..297C, 1998ApJ...507..327P, 2003A&A...409..581P}.  It would be an interesting exercise to repeat our analysis using a different calculation, such as DRAGON by \cite{DRAGONRef} \citep{Evoli:2008dv, DiBernardo:2009ku}, USINE by \cite{USINERef}, PPPC4DMID by \cite{PPPC4DMID} or the code of \cite{2003ICRC....4.1985B} to confirm these possibilities.\footnote{We thank the referee of our manuscript for alerting us to some of the possibilities discussed, and for suggesting some of the references, in this paragraph.}

Assuming that the propagation model satisfactorily describes physics over the Galaxy the next reasonable thing is to suspect local effects modifying the electron-positron distribution \citep{PesceRollins:2009tu}.  Further suspicion falls on the lack of sources included in the calculation \citep{DiBernardo:2010is, deVega:2010wj, Blum:2010nx, Frandsen:2010mr}.  Possible new sources of cosmic rays to account for the anomaly have been proposed in two major categories.  The first category is known astrophysical objects with unknown or uncertain parameters \citep{Lavalle:2010sf}.  These could be supernova remnants, pulsars, various objects in the Galactic centre, etc. \citep{Kawanaka:2010uj, Kashiyama:2010ui, Pato:2010im, Yuan:2011ys, Guo:2011dy, DiBernardo:2011wm}.  Finally, more exotic explanations call for new astronomical and/or particle physics phenomena, such as dark matter \citep{Pieri:2009je, Abidin:2010ea, Josan:2010vn, Cheng:2010mw, Ko:2010at, Cirelli:2010xx, Cholis:2010px, PalomaresRuiz:2010uu, Anderson:2010hh, Zaharijas:2010ca, Yang:2010zzd, Borriello:2010qh, Kajiyama:2010sb, Finkbeiner:2010sm, Buckley:2010ve, Kyae:2010sh, Logan:2010nw, Hutsi:2010ai, Feldman:2010wy, Arina:2010wv, Cholis:2010xb, Chen:2010yi, Lineros:2010ik, Dugger:2010ys, Vincent:2010kv, Mohanty:2010es, Bell:2010ei, Kang:2010mh, Haba:2010ag, Cline:2010fq, Ishiwata:2010am, Barger:2010mc, Kang:2010ha, Carone:2010ha, Cirelli:2010nh, 
Masina:2011ew, Porter:2011nv, Hutsi:2011vx, Sanchez:2011mf, Zavala:2011tt, Ke:2011xw, Zhu:2011dz, Bell:2011eu}.

Possible deviations from the predicted background can occur for energies above 100 GeV, as electron propagation is limited by energy losses via inverse Compton scattering of interstellar dust and CMB light, and synchrotron radiation due to Galactic magnetic field. This results in a relatively short lifetime and a rapidly decreasing intensity of the cosmic ray, as energy increases. Hence a large fraction of the detected electrons and positrons above 100 GeV are hypothesised to come from individual nearby sources that are within a few kilo-parsecs of the Earth \citep{Delahaye:2009gd, Grasso:2009ma}. Random fluctuations in the injection spectrum and the spatial distribution of those nearby sources produce significant differences in the most energetic part of the observed electron and positron spectrum. This can be the indication of new physics either from an astrophysical object(s) or dark matter. 

\begin{figure}
\begin{center}
\includegraphics[width=0.47\textwidth]{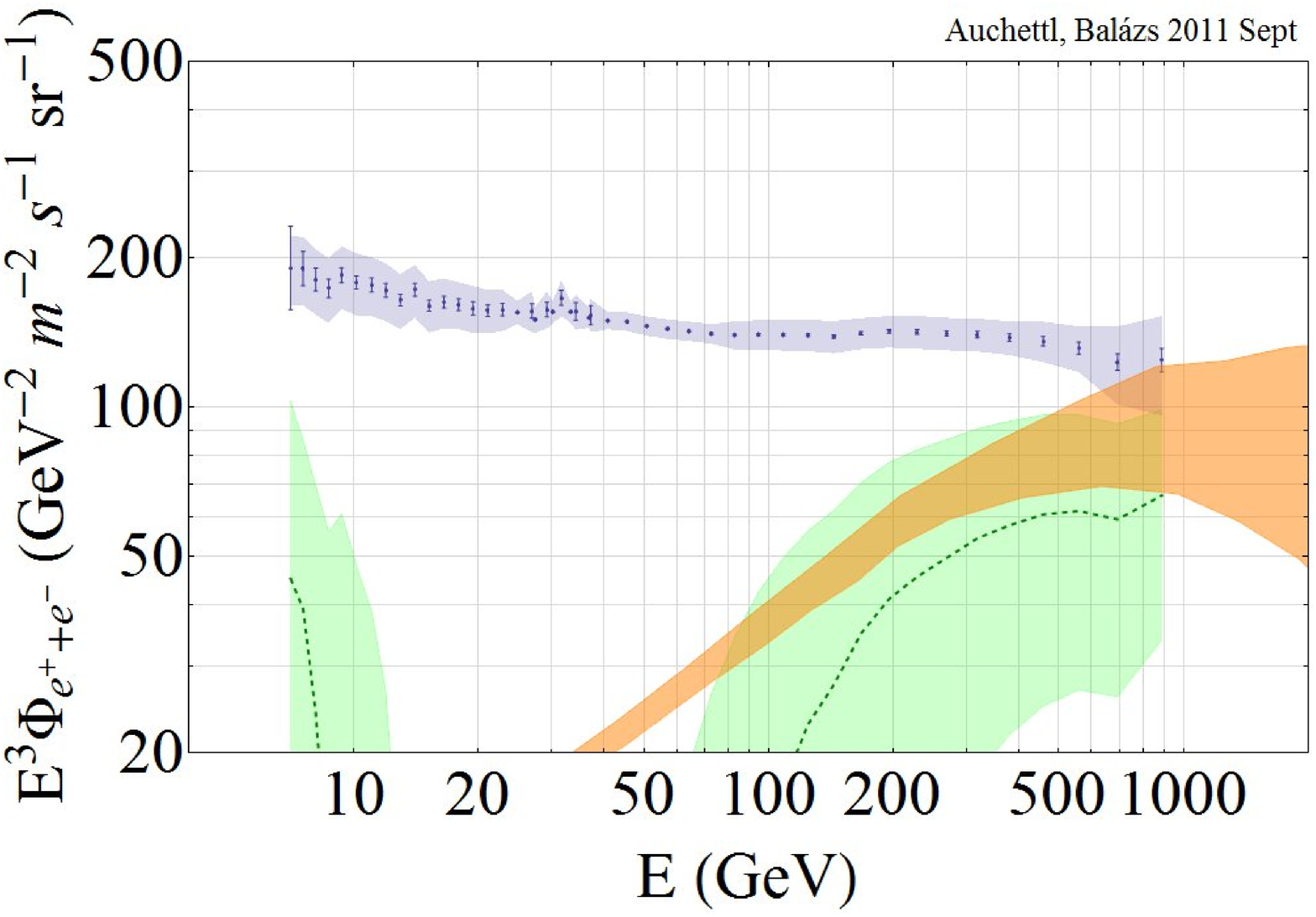}\vspace{3mm}
\includegraphics[width=0.47\textwidth]{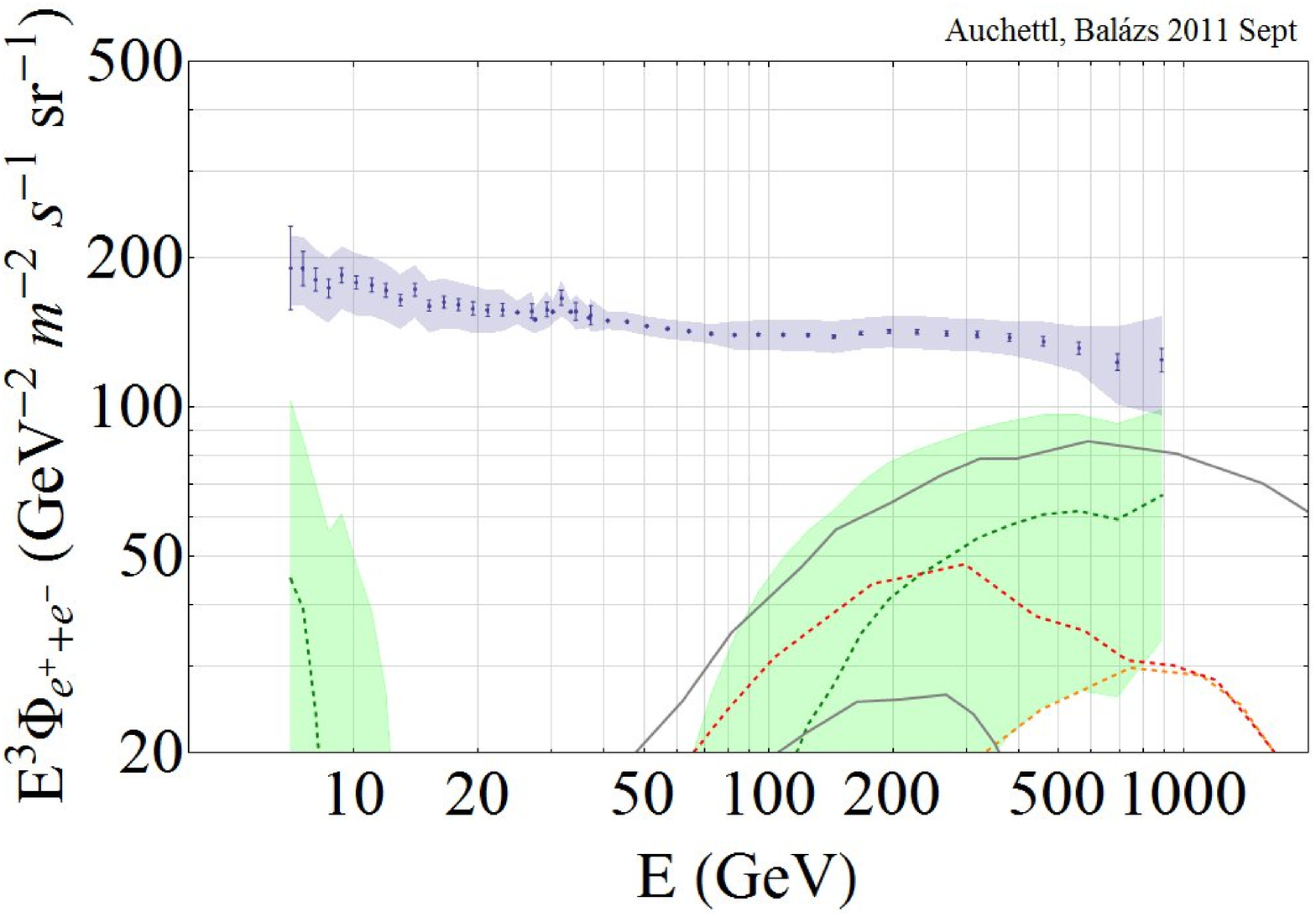}
\includegraphics[width=0.47\textwidth]{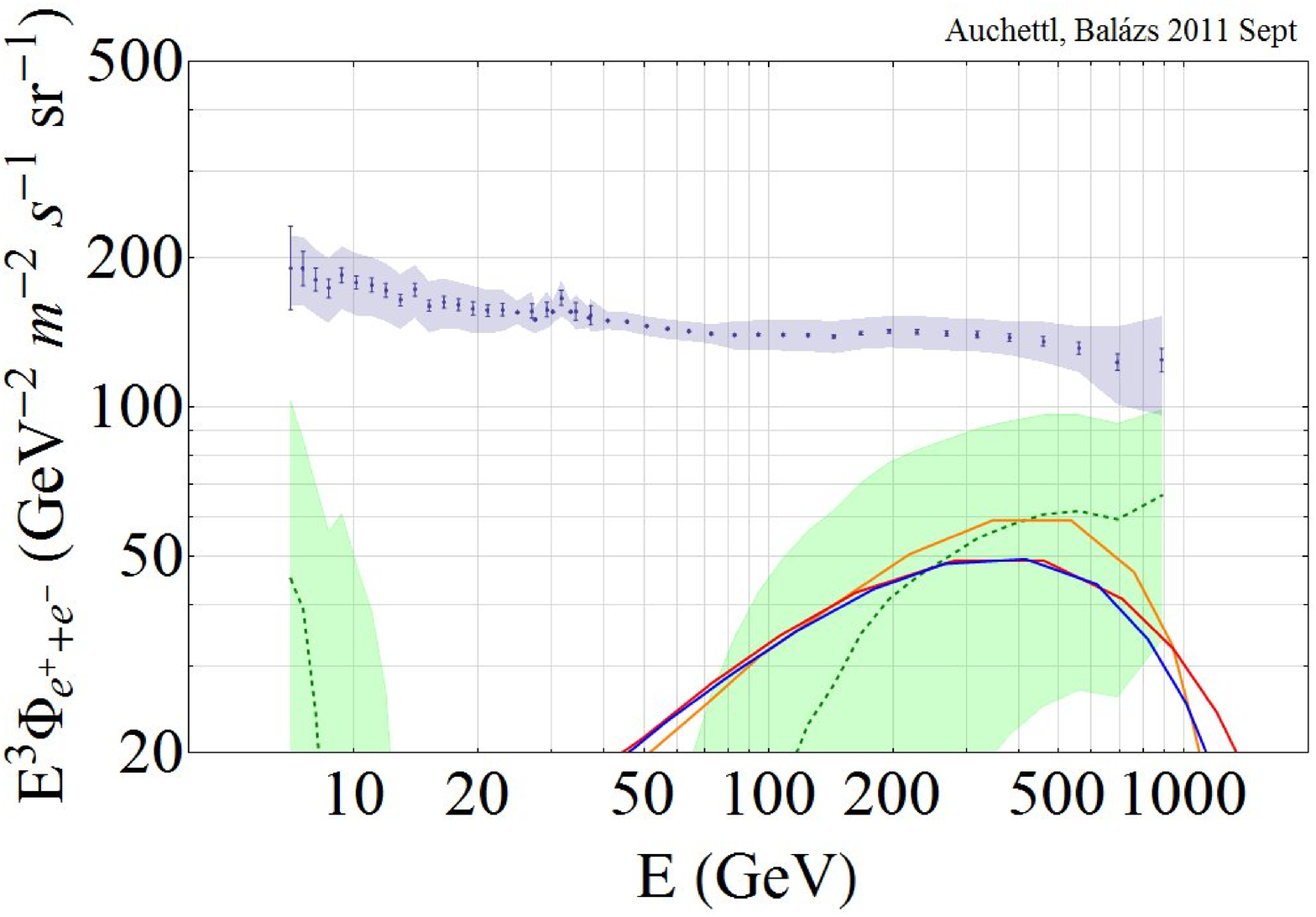}
\includegraphics[width=0.47\textwidth]{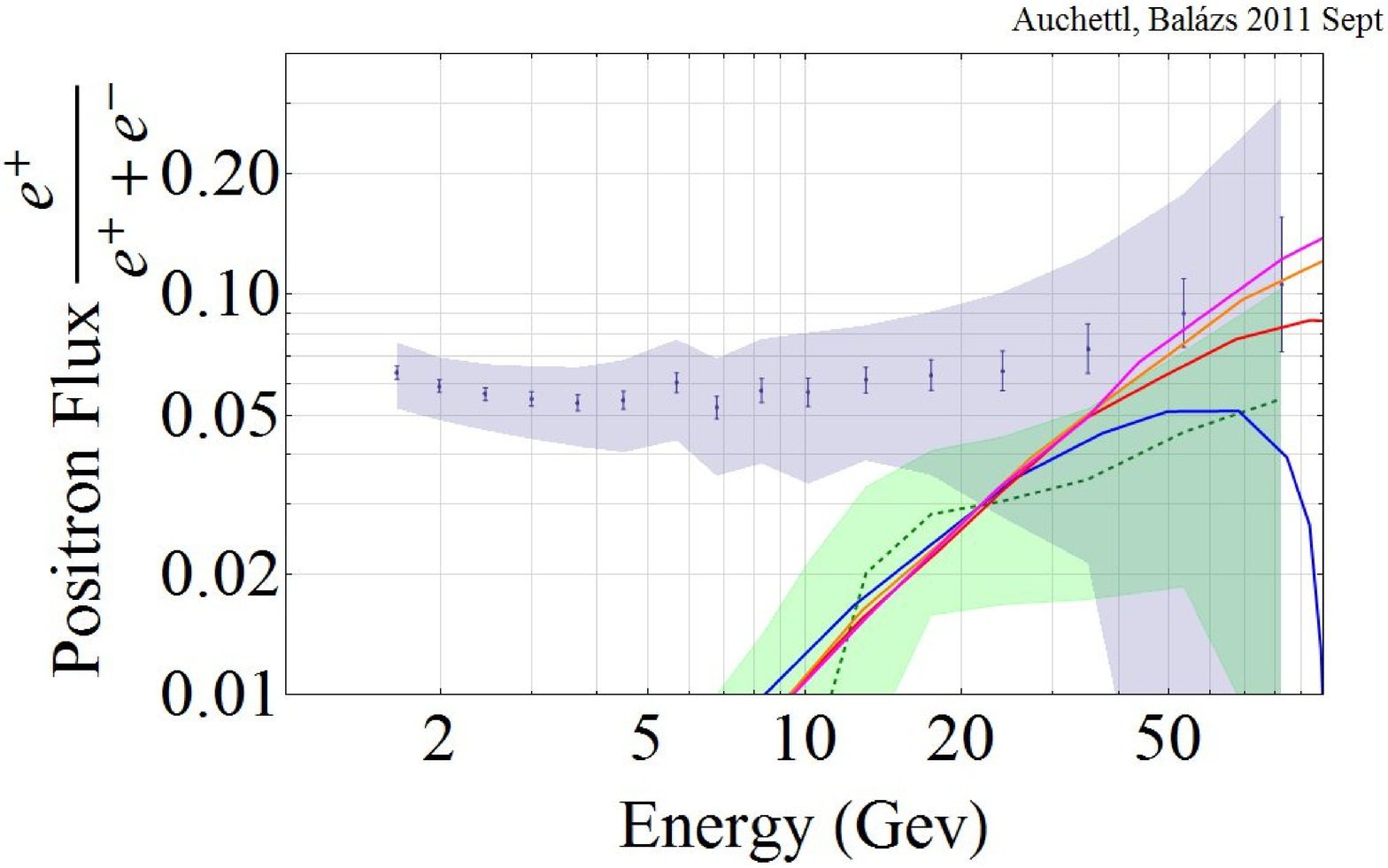}
\end{center}
\caption{Comparison of the signal extracted in this work to potential explanations of the electron-positron cosmic ray anomaly.  The various theoretical predictions come from \cite{Ahlers:2009ae}, \cite{Grasso:2009ma}, \cite{Bergstrom:2009fa} and \cite{Cholis:2008qq}.  Presently the comparison is fairly inconclusive but with more data it will be possible to shrink the uncertainty in the determination of the signal.  Then various suggestions can be confirmed or ruled out.}
\label{fig:Sources}
\end{figure}

Whatever the source of the anomaly is, if the size of the anomaly can be isolated then the source will have to match that size.  Fig. \ref{fig:Sources} compares our extracted signal to a few attempts to match the anomaly that we randomly selected from the recent literature.  The first frame shows the prediction of \cite{Ahlers:2009ae} for unaccounted energetic electrons and positrons produced by supernova remnants.  The top right frame features contributions from additional electron-positron primary sources (nearby pulsars or particle dark matter annihilation) calculated by \cite{Grasso:2009ma}.  The bottom left frame contains predictions of \cite{Bergstrom:2009fa} for anomalous electron-positron sources from dark matter annihilations.  Similar to this, dark matter annihilation contributions suggested by \cite{Cholis:2008qq} are shown in the last frame.  

The contributions of various new sources typically come with their own (theoretical) uncertainties.  In some of the cases this uncertainty is unknown thus it is hard to draw any conclusion by comparing these speculations to our isolated signal.  In the cases where the theoretical uncertainty is known, presently it tends to be large enough to prevent us from judging the validity of the given explanation.  Nevertheless, based on the present amount of information, we can already select a few scenarios that are more favored than some others.  By adding more data to our analysis it is possible to shrink the uncertainty of the signal.  Similarly, in most cases, the theoretical model of a given new source can be constrained further producing a narrower prediction.  With time, more data and more precise calculations the various suggestions of the cosmic electron-positron anomaly can be ruled out or confirmed.

\section{Conclusions}

We subjected a wide range of cosmic ray observations to a Bayesian likelihood analysis, motivated by the possibility of new physics contributing to the measurements of PAMELA and Fermi-LAT.  In the context of the propagation model coded in GalProp, we found a significant tension between the electron-positron related data and the rest of the cosmic ray fluxes.  This tension can be interpreted as the failure of the model to describe all the data simultaneously or as the effect of a missing source component.

Since the PAMELA and Fermi-LAT data are suspected to contain a component unaccounted for in GalProp, we extracted the preferred values of the cosmic ray propagation parameters from the non-electron-positron related measurements.  Based on these parameter values we calculated background predictions, with uncertainties, for PAMELA and Fermi-LAT.  We found a deviation between the PAMELA and Fermi-LAT data and the predicted background even when uncertainties, including systematics, were taken into account.  Interpreting this as an indication of new physics we subtracted the background from the data isolating the size of the anomalous component.  

The signal of new physics in the electron+positron spectrum was found to be non-vanishing within the calculated uncertainties.  Thus the use of 219 cosmic ray spectral data points within the Bayesian framework allowed us to confirm the existence of new physics effects in the electron+positron flux in a model independent fashion.  Using the statistical techniques we were able to extract the size, shape and uncertainty of the anomalous contribution in the electron+positron cosmic ray spectrum.  We briefly compared the extracted signal to some theoretical results predicting such an anomaly.

\acknowledgments

The authors are indebted to R. Cotta, A. Donea, J. Lazendic-Galloway, Y. Levin, A. Mazumdar, N. Sahu and T. Porter for invaluable discussions on various aspects of cosmic ray physics and likelihood analysis.  KA is thankful to P. Chan for help with issues of parallel computing, to M. Jasperse for assistance in Mathematica programming, to J. Lazendic-Galloway, T. Porter and to A. Vladimirov for help with GalProp.
This research was funded in part by the Australian Research Council under Project ID DP0877916 and in part by the Project of Knowledge Innovation Program (PKIP) of Chinese Academy of Sciences, Grant No. KJCX2.YW.W10.  The use of Monash University Sun Grid, a high-performance computing facility, is also gratefully acknowledged.

\bibliographystyle{apj}
\bibliography{paper}

\begin{thebibliography}{201}
\expandafter\ifx\csname natexlab\endcsname\relax\def\natexlab#1{#1}\fi

\bibitem[{Abdo {et~al.}(2009)}]{Abdo:2009zk}
Abdo, A.~A., {et~al.} 2009, Phys. Rev. Lett., 102, 181101

\bibitem[{Abidin {et~al.}(2010)Abidin, Afanasev, \& Carlson}]{Abidin:2010ea}
Abidin, Z., Afanasev, A., \& Carlson, C.~E. 2010, arXiv:1006.5444

\bibitem[{Ackermann {et~al.}(2010)}]{Ackermann:2010ij}
Ackermann, M., {et~al.} 2010, Phys. Rev., D82, 092004

\bibitem[{Adriani {et~al.}(2010{\natexlab{a}})Adriani, Barbarino, Bazilevskaya,
  Bellotti, Boezio, {et~al.}}]{Adriani:2010ib}
Adriani, O., Barbarino, G., Bazilevskaya, G., Bellotti, R., Boezio, M.,
  {et~al.} 2010{\natexlab{a}}, Astropart.Phys., 34, 1

\bibitem[{Adriani {et~al.}(2009)}]{Adriani:2008zr}
Adriani, O., {et~al.} 2009, Nature, 458, 607

\bibitem[{Adriani {et~al.}(2010{\natexlab{b}})}]{Adriani:2010rc}
---. 2010{\natexlab{b}}, Phys. Rev. Lett., 105, 121101

\bibitem[{Adriani {et~al.}(2011)}]{Adriani:2011xv}
---. 2011, Phys. Rev. Lett., 106, 201101

\bibitem[{Aguilar {et~al.}(2002)}]{Aguilar:2002ad}
Aguilar, M., {et~al.} 2002, Phys. Rept., 366, 331

\bibitem[{Aharonian {et~al.}(2011)Aharonian, Bykov, Parizot, Ptuskin, \&
  Watson}]{Aharonian:2011da}
Aharonian, F., Bykov, A., Parizot, E., Ptuskin, V., \& Watson, A. 2011,
  arXiv:1105.0131

\bibitem[{Aharonian {et~al.}(2008)}]{Aharonian:2008aa}
Aharonian, F., {et~al.} 2008, Phys. Rev. Lett., 101, 261104

\bibitem[{Aharonian {et~al.}(2009)}]{Aharonian:2009ah}
---. 2009, Astron. Astrophys., 508, 561

\bibitem[{Ahlers {et~al.}(2009)Ahlers, Mertsch, \& Sarkar}]{Ahlers:2009ae}
Ahlers, M., Mertsch, P., \& Sarkar, S. 2009, Phys. Rev., D80, 123017

\bibitem[{Ahn {et~al.}(2008)}]{Ahn:2008my}
Ahn, H.~S., {et~al.} 2008, Astropart. Phys., 30, 133

\bibitem[{Alcaraz {et~al.}(2000)}]{Alcaraz:2000bf}
Alcaraz, J., {et~al.} 2000, Phys. Lett., B484, 10

\bibitem[{Allahverdi {et~al.}(2009)Allahverdi, Dutta, Richardson-McDaniel, \&
  Santoso}]{Allahverdi:2008jm}
Allahverdi, R., Dutta, B., Richardson-McDaniel, K., \& Santoso, Y. 2009, Phys.
  Rev., D79, 075005

\bibitem[{Anderson(2010)}]{Anderson:2010hh}
Anderson, B. 2010, arXiv:1012.0863

\bibitem[{Arakida \& Kuramata(2011)}]{Arakida:2010vz}
Arakida, H., \& Kuramata, S. 2011, Int.J.Mod.Phys., A26, 911

\bibitem[{Arina {et~al.}(2010)Arina, Josse-Michaux, \& Sahu}]{Arina:2010wv}
Arina, C., Josse-Michaux, F.-X., \& Sahu, N. 2010, Phys.Lett., B691, 219

\bibitem[{Arkani-Hamed {et~al.}(2009)Arkani-Hamed, Finkbeiner, Slatyer, \&
  Weiner}]{ArkaniHamed:2008qn}
Arkani-Hamed, N., Finkbeiner, D.~P., Slatyer, T.~R., \& Weiner, N. 2009, Phys.
  Rev., D79, 015014

\bibitem[{Arvanitaki {et~al.}(2009)}]{Arvanitaki:2009yb}
Arvanitaki, A., {et~al.} 2009, Phys. Rev., D80, 055011

\bibitem[{Backovic \& Ralston(2010)}]{Backovic:2009rw}
Backovic, M., \& Ralston, J.~P. 2010, Phys. Rev., D81, 056002

\bibitem[{Bai {et~al.}(2009)Bai, Carena, \& Lykken}]{Bai:2009ka}
Bai, Y., Carena, M., \& Lykken, J. 2009, Phys. Rev., D80, 055004

\bibitem[{Baltz \& Edsj\"o(1998)}]{PhysRevD.59.023511}
Baltz, E.~A., \& Edsj\"o, J. 1998, Phys. Rev. D, 59, 023511

\bibitem[{Barger {et~al.}(2009)Barger, Gao, Keung, Marfatia, \&
  Shaughnessy}]{Barger:2009yt}
Barger, V., Gao, Y., Keung, W.~Y., Marfatia, D., \& Shaughnessy, G. 2009, Phys.
  Lett., B678, 283

\bibitem[{Barger {et~al.}(2010)Barger, Gao, McCaskey, \&
  Shaughnessy}]{Barger:2010mc}
Barger, V., Gao, Y., McCaskey, M., \& Shaughnessy, G. 2010, Phys.Rev., D82,
  095011

\bibitem[{Barwick {et~al.}(1997)}]{Barwick:1997ig}
Barwick, S.~W., {et~al.} 1997, Astrophys. J., 482, L191

\bibitem[{{Beach} {et~al.}(2001){Beach}, {Beatty}, {Bhattacharyya}, {Bower},
  {Coutu}, {Duvernois}, {Labrador}, {McKee}, {Minnick}, {M{\"u}ller}, {Musser},
  {Nutter}, {Schubnell}, {Swordy}, {Tarl{\'e}}, \&
  {Tomasch}}]{2001PhRvL..87A1101B}
{Beach}, A.~S., {et~al.} 2001, Physical Review Letters, 87, A261101+

\bibitem[{Beatty {et~al.}(2004)}]{Beatty:2004cy}
Beatty, J.~J., {et~al.} 2004, Phys. Rev. Lett., 93, 241102

\bibitem[{Bell {et~al.}(2011{\natexlab{a}})Bell, Dent, Jacques, \&
  Weiler}]{Bell:2011eu}
Bell, N.~F., Dent, J.~B., Jacques, T.~D., \& Weiler, T.~J. 2011{\natexlab{a}},
  arXiv:1101.3357

\bibitem[{Bell {et~al.}(2011{\natexlab{b}})Bell, Dent, Jacques, \&
  Weiler}]{Bell:2010ei}
---. 2011{\natexlab{b}}, Phys.Rev., D83, 013001

\bibitem[{Belotsky {et~al.}(2008)Belotsky, Fargion, Khlopov, \&
  Konoplich}]{Belotsky:2004st}
Belotsky, K., Fargion, D., Khlopov, M., \& Konoplich, R. 2008, Phys.Atom.Nucl.,
  71, 147

\bibitem[{Bergstrom {et~al.}(2009)Bergstrom, Edsjo, \&
  Zaharijas}]{Bergstrom:2009fa}
Bergstrom, L., Edsjo, J., \& Zaharijas, G. 2009, Phys. Rev. Lett., 103, 031103

\bibitem[{Bi {et~al.}(2009)Bi, He, \& Yuan}]{Bi:2009uj}
Bi, X.-J., He, X.-G., \& Yuan, Q. 2009, Phys. Lett., B678, 168

\bibitem[{{Blandford} \& {Eichler}(1987)}]{1987PhR...154....1B}
{Blandford}, R., \& {Eichler}, D. 1987, \physrep, 154, 1

\bibitem[{Blasi(2009)}]{Blasi:2009hv}
Blasi, P. 2009, Phys. Rev. Lett., 103, 051104

\bibitem[{Blum(2010)}]{Blum:2010nx}
Blum, K. 2010, arXiv:1010.2836

\bibitem[{Boezio {et~al.}(2001)}]{Boezio:2001ac}
Boezio, M., {et~al.} 2001, Astrophys. J., 561, 787

\bibitem[{Borriello {et~al.}(2010)Borriello, Maccione, \&
  Cuoco}]{Borriello:2010qh}
Borriello, E., Maccione, L., \& Cuoco, A. 2010, arXiv:1012.0041

\bibitem[{Brun {et~al.}(2009)Brun, Delahaye, Diemand, Profumo, \&
  Salati}]{Brun:2009aj}
Brun, P., Delahaye, T., Diemand, J., Profumo, S., \& Salati, P. 2009, Phys.
  Rev., D80, 035023

\bibitem[{Buchmuller {et~al.}(2009)Buchmuller, Ibarra, Shindou, Takayama, \&
  Tran}]{Buchmuller:2009xv}
Buchmuller, W., Ibarra, A., Shindou, T., Takayama, F., \& Tran, D. 2009, JCAP,
  0909, 021

\bibitem[{Buckley {et~al.}(2010)Buckley, Hooper, \& Tait}]{Buckley:2010ve}
Buckley, M.~R., Hooper, D., \& Tait, T.~M. 2010, arXiv:1011.1499

\bibitem[{{Buesching} {et~al.}(2003){Buesching}, {Kopp}, {Pohl}, \&
  {Shlickeiser}}]{2003ICRC....4.1985B}
{Buesching}, I., {Kopp}, A., {Pohl}, M., \& {Shlickeiser}, R. 2003, in
  International Cosmic Ray Conference, Vol.~4, International Cosmic Ray
  Conference, 1985--+

\bibitem[{{Burger}(2004)}]{2004EPJC...33S.941B2}
{Burger}, J. 2004, European Physical Journal C, 33, 941

\bibitem[{Calmet \& Majee(2009)}]{Calmet:2009uz}
Calmet, X., \& Majee, S.~K. 2009, Phys. Lett., B679, 267

\bibitem[{Carone {et~al.}(2010)Carone, Erlich, \& Primulando}]{Carone:2010ha}
Carone, C.~D., Erlich, J., \& Primulando, R. 2010, Phys.Rev., D82, 055028

\bibitem[{Chen(2009)}]{Chen:2009zpa}
Chen, C.-H. 2009, arXiv:0905.3425

\bibitem[{Chen {et~al.}(2009)Chen, Geng, \& Zhuridov}]{Chen:2009gd}
Chen, C.-H., Geng, C.-Q., \& Zhuridov, D.~V. 2009, JCAP, 0910, 001

\bibitem[{Chen {et~al.}(2010)Chen, Geng, \& Zhuridov}]{Chen:2009mj}
---. 2010, Eur. Phys. J., C67, 479

\bibitem[{Chen {et~al.}(2011)Chen, Feldman, Liu, Nath, \& Peim}]{Chen:2010yi}
Chen, N., Feldman, D., Liu, Z., Nath, P., \& Peim, G. 2011, Phys.Rev., D83,
  023506

\bibitem[{Cheng {et~al.}(2011)Cheng, Huang, Low, \& Menon}]{Cheng:2010mw}
Cheng, H.-C., Huang, W.-C., Low, I., \& Menon, A. 2011, JHEP, 1103, 019

\bibitem[{{Childers} \& {Duvernois}(2008)}]{2008ICRC....2..183C}
{Childers}, J.~T., \& {Duvernois}, M.~A. 2008, in International Cosmic Ray
  Conference, Vol.~2, International Cosmic Ray Conference, 183--186

\bibitem[{Choi \& Yaguna(2010)}]{Choi:2009qc}
Choi, K.-Y., \& Yaguna, C.~E. 2010, Phys. Rev., D81, 023502

\bibitem[{Cholis(2010)}]{Cholis:2010xb}
Cholis, I. 2010, arXiv:1007.1160

\bibitem[{Cholis {et~al.}(2009)Cholis, Finkbeiner, Goodenough, \&
  Weiner}]{Cholis:2008qq}
Cholis, I., Finkbeiner, D.~P., Goodenough, L., \& Weiner, N. 2009, JCAP, 0912,
  007

\bibitem[{Cholis \& Goodenough(2010)}]{Cholis:2010px}
Cholis, I., \& Goodenough, L. 2010, JCAP, 1009, 010

\bibitem[{Ciafaloni {et~al.}(2011)Ciafaloni, Cirelli, Comelli, De~Simone,
  Riotto, {et~al.}}]{Ciafaloni:2011sa}
Ciafaloni, P., Cirelli, M., Comelli, D., De~Simone, A., Riotto, A., {et~al.}
  2011, JCAP, 1106, 018

\bibitem[{Cirelli \& Cline(2010)}]{Cirelli:2010nh}
Cirelli, M., \& Cline, J.~M. 2010, Phys.Rev., D82, 023503

\bibitem[{Cirelli {et~al.}(2011{\natexlab{a}})Cirelli, Corcella, Hektor, Hutsi,
  Kadastik, {et~al.}}]{PPPC4DMID}
Cirelli, M., Corcella, G., Hektor, A., Hutsi, G., Kadastik, M., {et~al.}
  2011{\natexlab{a}}, JCAP, 1103, 051

\bibitem[{Cirelli {et~al.}(2011{\natexlab{b}})Cirelli, Corcella, Hektor, Hutsi,
  Kadastik, {et~al.}}]{Cirelli:2010xx}
---. 2011{\natexlab{b}}, JCAP, 1103, 051

\bibitem[{Cirelli {et~al.}(2009)Cirelli, Kadastik, Raidal, \&
  Strumia}]{Cirelli:2008pk}
Cirelli, M., Kadastik, M., Raidal, M., \& Strumia, A. 2009, Nucl. Phys., B813,
  1

\bibitem[{Cline(2010)}]{Cline:2010fq}
Cline, J.~M. 2010, arXiv:1005.5001

\bibitem[{Cotta {et~al.}(2011)Cotta, Conley, Gainer, Hewett, \&
  Rizzo}]{Cotta:2010ej}
Cotta, R.~C., Conley, J.~A., Gainer, J.~S., Hewett, J.~L., \& Rizzo, T.~G.
  2011, JHEP, 01, 064

\bibitem[{Cowsik \& Burch(2009)}]{Cowsik:2009ga}
Cowsik, R., \& Burch, B. 2009, arXiv:0905.2136

\bibitem[{{Cowsik} \& {Lee}(1979)}]{1979ApJ...228..297C}
{Cowsik}, R., \& {Lee}, M.~A. 1979, \apj, 228, 297

\bibitem[{Dado \& Dar(2009)}]{Dado:2009ux}
Dado, S., \& Dar, A. 2009, arXiv:0903.0165

\bibitem[{{Davis} {et~al.}(2000){Davis}, {Mewaldt}, {Binns}, {Christian},
  {Cummings}, {George}, {Hink}, {Leske}, {von Rosenvinge}, {Wiedenbeck}, \&
  {Yanasak}}]{2000AIPC..528..421D}
{Davis}, A.~J., {et~al.} 2000, in American Institute of Physics Conference
  Series, Vol. 528, Acceleration and Transport of Energetic Particles Observed
  in the Heliosphere, ed. {R.~A.~Mewaldt, J.~R.~Jokipii, M.~A.~Lee,
  E.~M{\"o}bius, \& T.~H.~Zurbuchen }, 421--424

\bibitem[{Davis {et~al.}(2000)Davis, Mewaldt, Binns, Christian, Cummings,
  George, Hink, Leske, von Rosenvinge, Wiedenbeck, \& Yanasak}]{davis:421}
Davis, A.~J., {et~al.} 2000, AIP Conference Proceedings, 528, 421

\bibitem[{De~Lope~Amigo {et~al.}(2009)De~Lope~Amigo, Cheung, Huang, \&
  Ng}]{DeLopeAmigo:2009dc}
De~Lope~Amigo, S., Cheung, W. M.-Y., Huang, Z., \& Ng, S.-P. 2009, JCAP, 0906,
  005

\bibitem[{de~Vega {et~al.}(2010)de~Vega, Falvella, \& Sanchez}]{deVega:2010wj}
de~Vega, H., Falvella, M., \& Sanchez, N. 2010, arXiv:1009.3494

\bibitem[{Delahaye {et~al.}(2011)Delahaye, Armand, Pohl, \&
  Salati}]{Timur:2011vv}
Delahaye, T., Armand, F., Pohl, M., \& Salati, P. 2011, arXiv:1102.0744

\bibitem[{Delahaye {et~al.}(2010)Delahaye, Lavalle, Lineros, Donato, \&
  Fornengo}]{Delahaye:2010ji}
Delahaye, T., Lavalle, J., Lineros, R., Donato, F., \& Fornengo, N. 2010,
  Astron. Astrophys., 524, A51

\bibitem[{Delahaye {et~al.}(2008)Delahaye, Lineros, Donato, Fornengo, \&
  Salati}]{Delahaye:2007fr}
Delahaye, T., Lineros, R., Donato, F., Fornengo, N., \& Salati, P. 2008, Phys.
  Rev., D77, 063527

\bibitem[{Delahaye {et~al.}(2009{\natexlab{a}})}]{Delahaye:2009gd}
Delahaye, T., {et~al.} 2009{\natexlab{a}}, arXiv:0905.2144

\bibitem[{Delahaye {et~al.}(2009{\natexlab{b}})}]{Delahaye:2008ua}
---. 2009{\natexlab{b}}, Astron. Astrophys., 501, 821

\bibitem[{Demir {et~al.}(2010)Demir, Everett, Frank, Selbuz, \&
  Turan}]{Demir:2009kc}
Demir, D.~A., Everett, L.~L., Frank, M., Selbuz, L., \& Turan, I. 2010, Phys.
  Rev., D81, 035019

\bibitem[{Dent {et~al.}(2010)Dent, Dutta, \& Scherrer}]{Dent:2009bv}
Dent, J.~B., Dutta, S., \& Scherrer, R.~J. 2010, Phys. Lett., B687, 275

\bibitem[{Di~Bernardo {et~al.}(2010)Di~Bernardo, Evoli, Gaggero, Grasso, \&
  Maccione}]{DiBernardo:2009ku}
Di~Bernardo, G., Evoli, C., Gaggero, D., Grasso, D., \& Maccione, L. 2010,
  Astropart. Phys., 34, 274

\bibitem[{Di~Bernardo {et~al.}(2011{\natexlab{a}})Di~Bernardo, Evoli, Gaggero,
  Grasso, Maccione, {et~al.}}]{DiBernardo:2011wm}
Di~Bernardo, G., Evoli, C., Gaggero, D., Grasso, D., Maccione, L., {et~al.}
  2011{\natexlab{a}}, arXiv:1101.1830

\bibitem[{Di~Bernardo {et~al.}(2011{\natexlab{b}})Di~Bernardo, Evoli, Gaggero,
  Grasso, Maccione, {et~al.}}]{DiBernardo:2010is}
---. 2011{\natexlab{b}}, Astropart.Phys., 34, 528

\bibitem[{Donato \& Serpico(2011)}]{Donato:2010vm}
Donato, F., \& Serpico, P.~D. 2011, Phys.Rev., D83, 023014

\bibitem[{Dugger {et~al.}(2010)Dugger, Jeltema, \& Profumo}]{Dugger:2010ys}
Dugger, L., Jeltema, T.~E., \& Profumo, S. 2010, JCAP, 1012, 015

\bibitem[{{Engelmann} {et~al.}(1990){Engelmann}, {Ferrando}, {Soutoul},
  {Goret}, \& {Juliusson}}]{1990AA...233...96E}
{Engelmann}, J.~J., {Ferrando}, P., {Soutoul}, A., {Goret}, P., \& {Juliusson},
  E. 1990, Astronomy and Astrophysics, 233, 96

\bibitem[{et~al.(2008)}]{PAMELABORONCARBON}
et~al., E.~M. 2008, in 21st European Cosmic Ray Symposium (ECRS 2008),
  Proceeding of 21st European Cosmic Ray Symposium, 396--401

\bibitem[{Evoli {et~al.}(2008)Evoli, Gaggero, Grasso, \&
  Maccione}]{Evoli:2008dv}
Evoli, C., Gaggero, D., Grasso, D., \& Maccione, L. 2008, JCAP, 0810, 018

\bibitem[{Fan {et~al.}(2010)Fan, Zhang, \& Chang}]{Fan:2010yq}
Fan, Y.-Z., Zhang, B., \& Chang, J. 2010, Int. J. Mod. Phys., D19, 2011

\bibitem[{Fargion {et~al.}(1999)Fargion, Golubkov, Khlopov, Konoplich, \&
  Mignani}]{Fargion:1999ss}
Fargion, D., Golubkov, Y., Khlopov, M.~Y., Konoplich, R., \& Mignani, R. 1999,
  JETP Lett., 69, 434, 10 pages, 4 PostScript figure, Latex2e

\bibitem[{Fargion {et~al.}(1995)Fargion, Khlopov, Konoplich, \&
  Mignani}]{PhysRevD.52.1828}
Fargion, D., Khlopov, M.~Y., Konoplich, R.~V., \& Mignani, R. 1995, Phys. Rev.
  D, 52, 1828

\bibitem[{Feldman {et~al.}(2009{\natexlab{a}})Feldman, Liu, \&
  Nath}]{Feldman:2008xs}
Feldman, D., Liu, Z., \& Nath, P. 2009{\natexlab{a}}, Phys. Rev., D79, 063509

\bibitem[{Feldman {et~al.}(2009{\natexlab{b}})Feldman, Liu, Nath, \&
  Nelson}]{Feldman:2009wv}
Feldman, D., Liu, Z., Nath, P., \& Nelson, B.~D. 2009{\natexlab{b}}, Phys.
  Rev., D80, 075001

\bibitem[{Feldman {et~al.}(2010)Feldman, Liu, Nath, \& Peim}]{Feldman:2010wy}
Feldman, D., Liu, Z., Nath, P., \& Peim, G. 2010, Phys.Rev., D81, 095017

\bibitem[{Feng {et~al.}(2010)Feng, Kaplinghat, \& Yu}]{Feng:2009hw}
Feng, J.~L., Kaplinghat, M., \& Yu, H.-B. 2010, Phys. Rev. Lett., 104, 151301

\bibitem[{Finkbeiner {et~al.}(2011)Finkbeiner, Goodenough, Slatyer,
  Vogelsberger, \& Weiner}]{Finkbeiner:2010sm}
Finkbeiner, D.~P., Goodenough, L., Slatyer, T.~R., Vogelsberger, M., \& Weiner,
  N. 2011, JCAP, 1105, 002

\bibitem[{Frandsen {et~al.}(2010)Frandsen, Masina, \&
  Sannino}]{Frandsen:2010mr}
Frandsen, M.~T., Masina, I., \& Sannino, F. 2010, arXiv:1011.0013

\bibitem[{Fujita {et~al.}(2009)}]{Fujita:2009wk}
Fujita, Y., {et~al.} 2009, Phys. Rev., D80, 063003

\bibitem[{Fukuoka {et~al.}(2009)Fukuoka, Kubo, \& Suematsu}]{Fukuoka:2009cu}
Fukuoka, H., Kubo, J., \& Suematsu, D. 2009, Phys. Lett., B678, 401

\bibitem[{{Garcia-Munoz} {et~al.}(1981){Garcia-Munoz}, {Guzik}, {Margolis},
  {Simpson}, \& {Wefel}}]{1981ICRC....9..195G}
{Garcia-Munoz}, M., {Guzik}, T.~G., {Margolis}, S.~H., {Simpson}, J.~A., \&
  {Wefel}, J.~P. 1981, in International Cosmic Ray Conference, Vol.~9,
  International Cosmic Ray Conference, 195--+

\bibitem[{Gast \& Schael(2009)}]{Gast:2009}
Gast, H., \& Schael, S. 2009, in 31st International Cosmic Ray Conference

\bibitem[{Ginzburg {et~al.}(1990)Ginzburg, Dogiel, Berezinsky, Bulanov, \&
  Ptuskin}]{Ginzburg:1990sk}
Ginzburg, V.L., e., Dogiel, V., Berezinsky, V., Bulanov, S., \& Ptuskin, V.
  1990, {Astrophysics of cosmic rays} ({North Holland})

\bibitem[{{Ginzburg} \& {Syrovatskii}(1964)}]{1964ocr..book.....G}
{Ginzburg}, V.~L., \& {Syrovatskii}, S.~I. 1964, {The Origin of Cosmic Rays}
  (Macmillan, New York)

\bibitem[{Golden {et~al.}(1994)}]{Golden:1992zm}
Golden, R.~L., {et~al.} 1994, Astrophys. J., 436, 769

\bibitem[{Grasso {et~al.}(2009)}]{Grasso:2009ma}
Grasso, D., {et~al.} 2009, Astropart. Phys., 32, 140

\bibitem[{Grimani {et~al.}(2002)}]{Grimani:2002yz}
Grimani, C., {et~al.} 2002, Astron. Astrophys., 392, 287

\bibitem[{Guo \& Wu(2009)}]{Guo:2009aj}
Guo, W.-L., \& Wu, Y.-L. 2009, Phys. Rev., D79, 055012

\bibitem[{Guo {et~al.}(2011)Guo, Feng, Yuan, Liu, \& Hu}]{Guo:2011dy}
Guo, Y., Feng, Z., Yuan, Q., Liu, C., \& Hu, H. 2011, arXiv:1101.5192

\bibitem[{Haba {et~al.}(2011)Haba, Kajiyama, Matsumoto, Okada, \&
  Yoshioka}]{Haba:2010ag}
Haba, N., Kajiyama, Y., Matsumoto, S., Okada, H., \& Yoshioka, K. 2011,
  Phys.Lett., B695, 476

\bibitem[{Hamaguchi {et~al.}(2009{\natexlab{a}})Hamaguchi, Nakaji, \&
  Nakamura}]{Hamaguchi:2009jb}
Hamaguchi, K., Nakaji, K., \& Nakamura, E. 2009{\natexlab{a}}, Phys. Lett.,
  B680, 172

\bibitem[{Hamaguchi {et~al.}(2009{\natexlab{b}})Hamaguchi, Nakamura, Shirai, \&
  Yanagida}]{Hamaguchi:2008rv}
Hamaguchi, K., Nakamura, E., Shirai, S., \& Yanagida, T.~T. 2009{\natexlab{b}},
  Phys. Lett., B674, 299

\bibitem[{{Hams} {et~al.}(2001){Hams}, {Barbier}, {Bremerich}, {Christian}, {de
  Nolfo}, {Geier}, {Goebel}, {Gupta}, {Hof}, {Menn}, {Mewaldt}, {Mitchell},
  {Schindler}, {Simon}, \& {Streitmatter}}]{2001ICRC....5.1655H}
{Hams}, T., {et~al.} 2001, in International Cosmic Ray Conference, Vol.~5,
  International Cosmic Ray Conference, 1655--+

\bibitem[{{Hareyama}(1999)}]{1999ICRC....3..105H}
{Hareyama}, M. 1999, in International Cosmic Ray Conference, Vol.~3,
  International Cosmic Ray Conference, 105--+

\bibitem[{Harnik \& Kribs(2009)}]{Harnik:2008uu}
Harnik, R., \& Kribs, G.~D. 2009, Phys. Rev., D79, 095007

\bibitem[{He(2009)}]{He:2009ra}
He, X.-G. 2009, Mod. Phys. Lett., A24, 2139

\bibitem[{{Hillas}(1984)}]{1984ARA&A..22..425H}
{Hillas}, A.~M. 1984, Ann. Rev. Astron. \& Astrophys., 22, 425

\bibitem[{Hisano {et~al.}(2005)Hisano, Matsumoto, Nojiri, \&
  Saito}]{Hisano:2004ds}
Hisano, J., Matsumoto, S., Nojiri, M.~M., \& Saito, O. 2005, Phys. Rev., D71,
  063528

\bibitem[{Hooper {et~al.}(2009{\natexlab{a}})Hooper, Blasi, \&
  Serpico}]{Hooper:2008kg}
Hooper, D., Blasi, P., \& Serpico, P.~D. 2009{\natexlab{a}}, JCAP, 0901, 025

\bibitem[{Hooper {et~al.}(2009{\natexlab{b}})Hooper, Stebbins, \&
  Zurek}]{Hooper:2008kv}
Hooper, D., Stebbins, A., \& Zurek, K.~M. 2009{\natexlab{b}}, Phys. Rev., D79,
  103513

\bibitem[{Hooper \& Tait(2009)}]{Hooper:2009gm}
Hooper, D., \& Tait, T. M.~P. 2009, Phys. Rev., D80, 055028

\bibitem[{Hu {et~al.}(2009)}]{Hu:2009bc}
Hu, H.-B., {et~al.} 2009, The Astrophysical Journal Letters, 700, L170

\bibitem[{Hutsi {et~al.}(2011)Hutsi, Chluba, Hektor, \& Raidal}]{Hutsi:2011vx}
Hutsi, G., Chluba, J., Hektor, A., \& Raidal, M. 2011, arXiv:1103.2766

\bibitem[{Hutsi {et~al.}(2010)Hutsi, Hektor, \& Raidal}]{Hutsi:2010ai}
Hutsi, G., Hektor, A., \& Raidal, M. 2010, JCAP, 1007, 008

\bibitem[{Ibarra \& Tran(2009)}]{Ibarra:2008jk}
Ibarra, A., \& Tran, D. 2009, JCAP, 0902, 021

\bibitem[{Ibarra {et~al.}(2010)Ibarra, Tran, \& Weniger}]{Ibarra:2009dr}
Ibarra, A., Tran, D., \& Weniger, C. 2010, JCAP, 1001, 009

\bibitem[{Ibe {et~al.}(2009)Ibe, Murayama, \& Yanagida}]{Ibe:2008ye}
Ibe, M., Murayama, H., \& Yanagida, T.~T. 2009, Phys. Rev., D79, 095009

\bibitem[{Ishiwata {et~al.}(2009)Ishiwata, Matsumoto, \&
  Moroi}]{Ishiwata:2008qy}
Ishiwata, K., Matsumoto, S., \& Moroi, T. 2009, Phys. Rev., D79, 043527

\bibitem[{Ishiwata {et~al.}(2010)Ishiwata, Matsumoto, \&
  Moroi}]{Ishiwata:2010am}
---. 2010, JHEP, 1012, 006

\bibitem[{Josan \& Green(2010)}]{Josan:2010vn}
Josan, A.~S., \& Green, A.~M. 2010, Phys.Rev., D82, 083527

\bibitem[{Kajiyama \& Okada(2011)}]{Kajiyama:2010sb}
Kajiyama, Y., \& Okada, H. 2011, Nucl.Phys., B848, 303

\bibitem[{Kane {et~al.}(2009)Kane, Lu, \& Watson}]{Kane:2009if}
Kane, G., Lu, R., \& Watson, S. 2009, Phys. Lett., B681, 151

\bibitem[{Kang \& Li(2011)}]{Kang:2010ha}
Kang, Z., \& Li, T. 2011, JHEP, 1102, 035

\bibitem[{Kang {et~al.}(2011)Kang, Li, Liu, Tong, \& Yang}]{Kang:2010mh}
Kang, Z., Li, T., Liu, T., Tong, C., \& Yang, J.~M. 2011, JCAP, 1101, 028

\bibitem[{Kashiyama {et~al.}(2011)Kashiyama, Ioka, \&
  Kawanaka}]{Kashiyama:2010ui}
Kashiyama, K., Ioka, K., \& Kawanaka, N. 2011, Phys.Rev., D83, 023002

\bibitem[{Katz {et~al.}(2009)Katz, Blum, \& Waxman}]{Katz:2009yd}
Katz, B., Blum, K., \& Waxman, E. 2009, Monthly Notices of the Royal
  Astronomical Society, 405, 1458

\bibitem[{Kawanaka {et~al.}(2011)Kawanaka, Ioka, Ohira, \&
  Kashiyama}]{Kawanaka:2010uj}
Kawanaka, N., Ioka, K., Ohira, Y., \& Kashiyama, K. 2011, Astrophys.J., 729, 93

\bibitem[{Ke {et~al.}(2011)Ke, Luo, Wang, \& Zhu}]{Ke:2011xw}
Ke, J., Luo, M., Wang, L., \& Zhu, G. 2011, Phys.Lett., B698, 44

\bibitem[{Ko \& Omura(2010)}]{Ko:2010at}
Ko, P., \& Omura, Y. 2010, arXiv:1012.4679

\bibitem[{{Krombel} \& {Wiedenbeck}(1988)}]{1988ApJ...328..940K}
{Krombel}, K.~E., \& {Wiedenbeck}, M.~E. 1988, Astrophys. J., 328, 940

\bibitem[{Kyae(2010)}]{Kyae:2010sh}
Kyae, B. 2010, J.Phys.Conf.Ser., 259, 012103

\bibitem[{Lavalle(2011)}]{Lavalle:2010sf}
Lavalle, J. 2011, arXiv1011.3063

\bibitem[{{Lezniak} \& {Webber}(1978)}]{1978ApJ...223..676L}
{Lezniak}, J.~A., \& {Webber}, W.~R. 1978, Astrophys. J., 223, 676

\bibitem[{Lin {et~al.}(2010)Lin, Finkbeiner, \& Dobler}]{Lin:2010fba}
Lin, T., Finkbeiner, D.~P., \& Dobler, G. 2010, Phys. Rev., D82, 023518

\bibitem[{Lineros(2010)}]{Lineros:2010ik}
Lineros, R.~A. 2010, J.Phys.Conf.Ser., 259, 012101

\bibitem[{Logan(2011)}]{Logan:2010nw}
Logan, H.~E. 2011, Phys.Rev., D83, 035022

\bibitem[{Maccione {et~al.}(2010)Maccione, Evoli, Gaggero, Di~Bernardo, \&
  Grasso}]{DRAGONRef}
Maccione, L., Evoli, C., Gaggero, D., Di~Bernardo, G., \& Grasso, D. 2010,
  {DRAGON: A public code to compute the propagation of high-energy Cosmic Rays
  in the Galaxy.}

\bibitem[{Malinin(2004)}]{Malinin:2004pw}
Malinin, A.~G. 2004, Phys. Atom. Nucl., 67, 2044

\bibitem[{Malyshev(2009)}]{Malyshev:2009zh}
Malyshev, D. 2009, JCAP, 0907, 038

\bibitem[{Malyshev {et~al.}(2009)Malyshev, Cholis, \&
  Gelfand}]{Malyshev:2009tw}
Malyshev, D., Cholis, I., \& Gelfand, J. 2009, Phys. Rev., D80, 063005

\bibitem[{March-Russell {et~al.}(2008)March-Russell, West, Cumberbatch, \&
  Hooper}]{MarchRussell:2008yu}
March-Russell, J., West, S.~M., Cumberbatch, D., \& Hooper, D. 2008, JHEP, 07,
  058

\bibitem[{Mardon {et~al.}(2009)Mardon, Nomura, \& Thaler}]{Mardon:2009gw}
Mardon, J., Nomura, Y., \& Thaler, J. 2009, Phys. Rev., D80, 035013

\bibitem[{Masina \& Sannino(2011)}]{Masina:2011ew}
Masina, I., \& Sannino, F. 2011, arXiv:1105.0302

\bibitem[{Maurin {et~al.}(2001)Maurin, Donato, Taillet, \&
  Salati}]{Maurin:2001sj}
Maurin, D., Donato, F., Taillet, R., \& Salati, P. 2001, Astrophys. J., 555,
  585

\bibitem[{{Maurin} {et~al.}(2010){Maurin}, {Putze}, \&
  {Derome}}]{2010A&A...516A..67M}
{Maurin}, D., {Putze}, A., \& {Derome}, L. 2010, Astronomy and Astrophysics,
  516, A67+

\bibitem[{Maurin {et~al.}(2011)Maurin, Putze, Derome, Taillet, Barao, Donato,
  Salati, \& Combet}]{USINERef}
Maurin, D., Putze, A., Derome, L., Taillet, R., Barao, F., Donato, F., Salati,
  P., \& Combet, C. 2011, {USINE - a galactic cosmic-ray propagation code.}

\bibitem[{Maurin {et~al.}(2002)Maurin, Taillet, \& Donato}]{Maurin:2002hw}
Maurin, D., Taillet, R., \& Donato, F. 2002, Astron. Astrophys., 394, 1039

\bibitem[{Mertsch(2010)}]{Mertsch:2010qf}
Mertsch, P. 2010, arXiv:1012.4239

\bibitem[{Mertsch \& Sarkar(2009)}]{Mertsch:2009ph}
Mertsch, P., \& Sarkar, S. 2009, Phys. Rev. Lett., 103, 081104

\bibitem[{Mitthumsiri(2011)}]{FermiPosFrac}
Mitthumsiri, W. 2011, in Fermi Symposium, Fermi Symposium

\bibitem[{Mohanty {et~al.}(2010)Mohanty, Rao, \& Roy}]{Mohanty:2010es}
Mohanty, S., Rao, S., \& Roy, D. 2010, arXiv:1009.5058

\bibitem[{Moskalenko {et~al.}(2002)Moskalenko, Strong, Ormes, \&
  Potgieter}]{Moskalenko:2001ya}
Moskalenko, I.~V., Strong, A.~W., Ormes, J.~F., \& Potgieter, M.~S. 2002,
  Astrophys.J., 565, 280

\bibitem[{Nakamura {et~al.}(2010)}]{Nakamura:2010zzi}
Nakamura, K., {et~al.} 2010, J.Phys.G, G37, 075021

\bibitem[{Nardi {et~al.}(2009)Nardi, Sannino, \& Strumia}]{Nardi:2008ix}
Nardi, E., Sannino, F., \& Strumia, A. 2009, JCAP, 0901, 043

\bibitem[{Okada \& Yamada(2009)}]{Okada:2009bz}
Okada, N., \& Yamada, T. 2009, Phys. Rev., D80, 075010

\bibitem[{Palomares-Ruiz \& Siegal-Gaskins(2010)}]{PalomaresRuiz:2010uu}
Palomares-Ruiz, S., \& Siegal-Gaskins, J.~M. 2010, arXiv:1012.2335

\bibitem[{Pato {et~al.}(2010)Pato, Lattanzi, \& Bertone}]{Pato:2010im}
Pato, M., Lattanzi, M., \& Bertone, G. 2010, JCAP, 1012, 020

\bibitem[{Perelstein \& Shakya(2010{\natexlab{a}})}]{Perelstein:2010gt}
Perelstein, M., \& Shakya, B. 2010{\natexlab{a}}, arXiv:1012.3772

\bibitem[{Perelstein \& Shakya(2010{\natexlab{b}})}]{Perelstein:2010fq}
---. 2010{\natexlab{b}}, Phys.Rev., D82, 043505

\bibitem[{Pesce-Rollins \& collaboration(2009)}]{PesceRollins:2009tu}
Pesce-Rollins, M., \& collaboration, f. t. F.~L. 2009, arXiv:0907.0387

\bibitem[{Pieri {et~al.}(2011)Pieri, Lavalle, Bertone, \&
  Branchini}]{Pieri:2009je}
Pieri, L., Lavalle, J., Bertone, G., \& Branchini, E. 2011, Phys.Rev., D83,
  023518

\bibitem[{{Pohl} \& {Esposito}(1998)}]{1998ApJ...507..327P}
{Pohl}, M., \& {Esposito}, J.~A. 1998, \apj, 507, 327

\bibitem[{{Pohl} {et~al.}(2003){Pohl}, {Perrot}, {Grenier}, \&
  {Digel}}]{2003A&A...409..581P}
{Pohl}, M., {Perrot}, C., {Grenier}, I., \& {Digel}, S. 2003, \aap, 409, 581

\bibitem[{Porter {et~al.}(2011)Porter, Johnson, \& Graham}]{Porter:2011nv}
Porter, T.~A., Johnson, R.~P., \& Graham, P.~W. 2011, Annual Review of
  Astronomy and Astrophysics, 49, null

\bibitem[{Prantzos {et~al.}(2010)Prantzos, Boehm, Bykov, Diehl, Ferriere,
  {et~al.}}]{Prantzos:2010wi}
Prantzos, N., Boehm, C., Bykov, A., Diehl, R., Ferriere, K., {et~al.} 2010,
  arXiv:1009.4620

\bibitem[{Profumo(2008)}]{Profumo:2008ms}
Profumo, S. 2008, arXiv:0812.4457

\bibitem[{Ptuskin {et~al.}(2006)Ptuskin, Moskalenko, Jones, Strong, \&
  Zirakashvili}]{Ptuskin:2005ax}
Ptuskin, V.~S., Moskalenko, I.~V., Jones, F.~C., Strong, A.~W., \&
  Zirakashvili, V.~N. 2006, Astrophys. J., 642, 902

\bibitem[{{Putze} {et~al.}(2010){Putze}, {Derome}, \&
  {Maurin}}]{2010A&A...516A..66P}
{Putze}, A., {Derome}, L., \& {Maurin}, D. 2010, Astronomy and Astrophysics,
  516, A66+

\bibitem[{Sanchez \& Holdom(2011)}]{Sanchez:2011mf}
Sanchez, C.~G., \& Holdom, B. 2011, arXiv:1103.1632

\bibitem[{Schlickeiser(2002)}]{Schlickeiser:2002pg}
Schlickeiser, R. 2002, {Cosmic ray astrophysics} ({Springer})

\bibitem[{{Seo} \& {Ptuskin}(1994)}]{1994ApJ...431..705S}
{Seo}, E.~S., \& {Ptuskin}, V.~S. 1994, Astrophys. J., 431, 705

\bibitem[{Serpico(2011)}]{Serpico:2011wg}
Serpico, P.~D. 2011, {Astrophysical models for the origin of the positron
  'excess'}

\bibitem[{Shaviv {et~al.}(2009)Shaviv, Nakar, \& Piran}]{Shaviv:2009bu}
Shaviv, N.~J., Nakar, E., \& Piran, T. 2009, Phys. Rev. Lett., 103, 111302

\bibitem[{Shirai {et~al.}(2009)Shirai, Takahashi, \& Yanagida}]{Shirai:2009fq}
Shirai, S., Takahashi, F., \& Yanagida, T.~T. 2009, Phys. Lett., B680, 485

\bibitem[{Shirai {et~al.}(2010)Shirai, Takahashi, \& Yanagida}]{Shirai:2009wi}
---. 2010, Prog. Theor. Phys., 122, 1277

\bibitem[{Stawarz {et~al.}(2010)Stawarz, Petrosian, \&
  Blandford}]{Stawarz:2009ig}
Stawarz, L., Petrosian, V., \& Blandford, R.~D. 2010, Astrophys.J., 710, 236

\bibitem[{Strong \& Moskalenko(1998)}]{Strong:1998pw}
Strong, A.~W., \& Moskalenko, I.~V. 1998, Astrophys. J., 509, 212

\bibitem[{Strong {et~al.}(2007)Strong, Moskalenko, \& Ptuskin}]{Strong:2007nh}
Strong, A.~W., Moskalenko, I.~V., \& Ptuskin, V.~S. 2007, Ann. Rev. Nucl. Part.
  Sci., 57, 285

\bibitem[{Strong {et~al.}(2004)Strong, Moskalenko, \& Reimer}]{Strong:2004de}
Strong, A.~W., Moskalenko, I.~V., \& Reimer, O. 2004, Astrophys. J., 613, 962

\bibitem[{Tawfik \& Saleh(2010)}]{Tawfik:2010qf}
Tawfik, A., \& Saleh, A. 2010, arXiv:1010.5390

\bibitem[{{Tierney} \& {Kadane}(1986)}]{Tierney1986}
{Tierney}, L., \& {Kadane}, J.~B. 1986, Journal of the American Statistical
  Association, 81, 82

\bibitem[{Torii {et~al.}(2008)}]{Torii:2008xu}
Torii, S., {et~al.} 2008, arXiv:0809.0760

\bibitem[{Trotta {et~al.}(2010)}]{Trotta:2010mx}
Trotta, R., {et~al.} 2010, The Astrophysical Journal, 729, 106

\bibitem[{{Usoskin} {et~al.}(2011){Usoskin}, {Bazilevskaya}, \&
  {Kovaltsov}}]{2011JGRA..11602104U}
{Usoskin}, I.~G., {Bazilevskaya}, G.~A., \& {Kovaltsov}, G.~A. 2011, Journal of
  Geophysical Research (Space Physics), 116, A02104

\bibitem[{Vincent {et~al.}(2010)Vincent, Xue, \& Cline}]{Vincent:2010kv}
Vincent, A.~C., Xue, W., \& Cline, J.~M. 2010, Phys.Rev., D82, 123519

\bibitem[{{Wiedenbeck} \& {Greiner}(1980)}]{1980ApJ...239L.139W}
{Wiedenbeck}, M.~E., \& {Greiner}, D.~E. 1980, \apjl, 239, L139

\bibitem[{{Yanasak} {et~al.}(2001){Yanasak}, {Wiedenbeck}, {Binns},
  {Christian}, {Cummings}, {Davis}, {George}, {Hink}, {Israel}, {Leske},
  {Lijowski}, {Mewaldt}, {Stone}, \& {von Rosenvinge}}]{2001AdSpR..27..727Y}
{Yanasak}, N.~E., {et~al.} 2001, Advances in Space Research, 27, 727

\bibitem[{Yang(2010)}]{Yang:2010zzd}
Yang, J.-M. 2010, Mod.Phys.Lett., A25, 976

\bibitem[{Yin {et~al.}(2009)}]{Yin:2008bs}
Yin, P.-f., {et~al.} 2009, Phys. Rev., D79, 023512

\bibitem[{Yuan {et~al.}(2011)Yuan, Zhang, \& Bi}]{Yuan:2011ys}
Yuan, Q., Zhang, B., \& Bi, X.-J. 2011, arXiv:1104.3357

\bibitem[{Yuksel {et~al.}(2009)Yuksel, Kistler, \& Stanev}]{Yuksel:2008rf}
Yuksel, H., Kistler, M.~D., \& Stanev, T. 2009, Phys. Rev. Lett., 103, 051101

\bibitem[{Zaharijas {et~al.}(2010)Zaharijas, Cuoco, Yang, \&
  Conrad}]{Zaharijas:2010ca}
Zaharijas, G., Cuoco, A., Yang, Z., \& Conrad, J. 2010, arXiv:1012.0588

\bibitem[{Zavala {et~al.}(2011)Zavala, Vogelsberger, Slatyer, Loeb, \&
  Springel}]{Zavala:2011tt}
Zavala, J., Vogelsberger, M., Slatyer, T.~R., Loeb, A., \& Springel, V. 2011,
  arXiv:1103.0776

\bibitem[{Zavala {et~al.}(2010)Zavala, Vogelsberger, \& White}]{Zavala:2009mi}
Zavala, J., Vogelsberger, M., \& White, S. D.~M. 2010, Phys. Rev., D81, 083502

\bibitem[{Zeldovich {et~al.}(1980)Zeldovich, Klypin, Khlopov, \&
  Chechetkin}]{Zeldovich:1980st}
Zeldovich, Y., Klypin, A., Khlopov, M., \& Chechetkin, V. 1980,
  Sov.J.Nucl.Phys., 31, 664

\bibitem[{Zhu(2011)}]{Zhu:2011dz}
Zhu, G. 2011, Phys.Rev., D83, 076011

\end{thebibliography}

%

\end{document}